\documentclass[11pt,a4paper]{article}
\pdfoutput=1
\usepackage{jheppub}
\usepackage[utf8]{inputenc}
\usepackage[english]{babel}
\makeatletter
\usepackage{amsmath,amssymb}
\usepackage{esint}
\usepackage{graphicx}
\usepackage{caption}
\usepackage{subcaption}
\usepackage{float}
\usepackage{mathtools}
\usepackage{xcolor}

\DeclareMathOperator{\tr}{Tr}

\title{Entanglement of a chiral fermion on the torus}
\author{David Blanco,}
\author{Alan Garbarz,}
\author{Guillem P\'erez-Nadal}
\affiliation{Universidad de Buenos Aires, Departamento de F\'isica and IFIBA - CONICET\\
1428 Buenos Aires, Argentina}
\emailAdd{dblanco@df.uba.ar, alan@df.uba.ar, guillem@df.uba.ar}

\abstract{
In this paper we present the detailed calculation of a new modular Hamiltonian, namely that of a chiral fermion on a circle at non-zero temperature. We provide explicit results for an arbitrary collection of intervals, which we discuss at length by checking against known results in different limits and by computing the associated modular flow. We also compute the entanglement entropy, and we obtain a simple expression for it which appears to be more manageable than those already existing in the literature.
}

\begin{document}

\maketitle

\section{Introduction}

In the last years, some ideas and results of quantum information theory have been applied in quantum field theory (QFT) with considerable success. For example, the strong subadditivity property of entanglement entropy has been used to show the irreversibility of the renormalization group flow in 2, 3 and 4 dimensions \cite{Casini:2006es,Casini:2012ei,Casini:2015woa,Casini:2017vbe}. The interplay with quantum information is also prominent in holography, where the Ryu-Takayanagi formula \cite{Ryu:2006bv} establishes a deep connection between entanglement and geometry. 

Quantum information measures are based on the reduced density matrix, or equivalently on what is essentially its logarithm, the modular Hamiltonian.
Among other applications, the knowledge of modular Hamiltonians was crucial for the proof of the averaged null energy condition \cite{Faulkner:2016mzt}, the derivation of quantum energy inequalities \cite{Blanco:2013lea,Blanco:2017akw} and the formulation of a well-defined version of the Bekenstein bound \cite{Casini:2008cr}. Modular Hamiltonians also played a key role in applications to holography, the most notable case probably being the derivation of the linearized Einstein equations in the bulk from entanglement properties of the boundary CFT \cite{Faulkner:2013ica,Lashkari:2013koa,Blanco:2018riw,Swingle:2014uza}.

There are only few cases where modular Hamiltonians have been computed. For any QFT in the vacuum, the modular Hamiltonian of the half-space (or its causal development, the Rindler wedge) is proportional to the generator of boosts \cite{Bisognano:1976za,Unruh:1976db}. In the case of a CFT, one can use this result to derive the modular Hamiltonian of a ball by applying a special conformal transformation \cite{Casini:2011kv}. Further restricting to 1+1 dimensions, conformal maps from the plane to the cylinder can be used to
obtain the modular Hamiltonian of an interval in other manifolds or states: a circle in the vacuum \cite{Cardy:2016fqc}, the line at non-zero temperature \cite{Hartman:2015apr}. Note that, since the boost generator is a local operator, all these modular Hamiltonians are local. In the case of free CFTs in the vacuum, some modular Hamiltonians are also known for multiple intervals: chiral fermion on the line \cite{Casini:2009vk} and the circle \cite{Wong:2013gua}, chiral scalar on the line \cite{Arias:2018tmw}. These modular Hamiltonians are not derivable from that of Rindler, and they turn out to be non-local. The non-locality disappears when restricting to a single interval, except in the case of the circle with Ramond boundary conditions, where the zero mode is responsible for the appearance of a non-local term even for a single interval.

In a recent paper \cite{Blanco:2019xwi} we computed a new modular Hamiltonian, namely that of a chiral fermion on a circle at non-zero temperature, i.e., on a torus in the Euclidean formalism (see \cite{Fries:2019ozf} for a related result). The difficulty of the torus is that it cannot be mapped conformally to the plane. However, there is a very natural way to go from the torus to the plane: simply unfold the torus. This is, in broad terms, the idea of our method. More precisely, for free theories in Gaussian states there is a general expression for the modular Hamiltonian in terms of the resolvent of a certain two-point function, which is viewed as an operator \cite{Casini:2009sr}. Our strategy to compute the resolvent is based on the method of images applied to the calculation of the Euclidean propagator, which enables us to effectively unfold the torus and map the problem to a similar problem on the plane. In this paper we provide full detail on this calculation, and extend
it by giving explicit results for multiple intervals. Interestingly, the modular Hamiltonian on the torus exhibits a completely non-local character, even for a single interval, regardless of the boundary conditions. We check the result by comparing it with known results in several limits, and we compute the modular flow, which inherits the non-locality of the modular Hamiltonian. Finally, we also use the resolvent to obtain a simple expression for the entanglement entropy of an arbitrary collection of intervals. The entanglement entropy on the torus had been previously computed in the form of a low-temperature and a high-temperature expansion \cite{Azeyanagi:2007bj,Herzog:2013py}; we find a simple exact result which appears to be more suitable for applications.

The paper is organized as follows. In section \ref{sect:1} we present the expressions that give the modular Hamiltonian and the entanglement entropy in terms of the resolvent. In section \ref{sect:2} we compute the resolvent on the torus by the method of images. In section \ref{sect:3} we use the resolvent to compute the modular Hamiltonian, which we check in different limits against known results in the literature, and which we use to compute the modular flow. In section \ref{sect:4} we obtain a closed formula for the entanglement entropy and study its behavior, also checking the result in several limits. We have also included an appendix on the Weierstrass functions, which we use extensively throughout the paper.

\section{Modular Hamiltonian and entropy from the resolvent}\label{sect:1}

Consider a chiral fermion $\psi$ on a circle of length $L$. The Hamiltonian is
\begin{equation}\label{ham}
    H=\pm i\int_{-L/2}^{L/2}dx\,\psi^\dagger\psi',
\end{equation}
where the sign depends on the chirality. Suppose that the field is in a thermal state with inverse temperature $\beta$. The purpose of this paper is to compute the modular Hamiltonian $H_V$ corresponding to a subset $V$ of the circle, which is related to the reduced density matrix $\rho_V$ by the equation
\begin{equation}
    \rho_V=\frac{e^{-H_V}}{\tr e^{-H_V}}.
\end{equation}
In other words, $H_V$ is the Hamiltonian that the system should have in order for $\rho_V$ to be thermal with unit temperature. We will also compute the entanglement entropy
\begin{equation}\label{EE}
    S_V=-\tr(\rho_V\log\rho_V)=\langle H_V\rangle+\log(\tr e^{-H_V}).
\end{equation}
Since the global state is Gaussian, correlation functions obey Wick's theorem. This is true everywhere in the circle, in particular within $V$, so the reduced density matrix is also Gaussian and hence the modular Hamiltonian has the form
\begin{equation}\label{mod}
    H_{V}=\int_V dx dy\,\psi^{\dagger}(x)K_{V}(x,y)\psi(y).
\end{equation}
The kernel $K_V$ is determined by the condition $G_V(x,y)=\tr(\rho_V\psi(x)\psi^\dagger(y))$, where $G_V$ is the two-point function $\langle\psi(x)\psi^\dagger(y)\rangle$ restricted to pairs of points in $V$. The result is \cite{peschel2003calculation,Casini:2009sr}
\begin{equation}\label{modG}
    K_{V}=-\log\left(G_{V}^{-1}-1\right),
\end{equation}
where both $K_V$ and $G_V$ are viewed as operators acting on functions on $V$. Note that $G_V$ is Hermitian and has its spectrum contained in the interval $(0,1)$, so $K_V$ is also Hermitian.
Substituting (\ref{mod}) and (\ref{modG}) into (\ref{EE}) one obtains \cite{Casini:2009sr}
\begin{equation}
    S_V=-\tr[(1-G_V)\log(1-G_V)+G_V\log G_V].
\end{equation}
The above two equations can be rewritten as
\begin{alignat}{2}
&K_V=-\int_{1/2}^\infty d\xi\left[R_V(\xi)+R_V(-\xi)\right]\label{modres}\\
&S_V=-\int_{1/2}^\infty d\xi\,\tr\left\{(\xi-1/2)\left[R_V(\xi)-R_V(-\xi)\right]-\frac{2\xi}{\xi+1/2}\right\},\label{EEres}   
\end{alignat}
where $R_V$ is the resolvent of $G_V$,
\begin{equation}
    R_V(\xi)=\frac{1}{G_V+\xi-1/2}.
\end{equation}
This can be easily checked by explicitly performing the integrals in (\ref{modres}) and (\ref{EEres}). Thus, the problem of computing the modular Hamiltonian and the entanglement entropy reduces to that of finding the resolvent of $G_V$.

\section{The resolvent}\label{sect:2}

\subsection{The method of images}

Our strategy for computing the resolvent is based on the method of images applied to the calculation of the Euclidean propagator $G$, which is defined by
\begin{alignat}{2}\label{eucl}
G(x,t;y,u)&=
    \theta(t-u)
    \langle e^{H(t-u)}\psi(x)e^{-H(t-u)}\psi^\dagger(y)\rangle\nonumber\\
    &-\theta(u-t)\langle \psi^\dagger(y)e^{H(t-u)}\psi(x)e^{-H(t-u)}\rangle 
\end{alignat}
for $t-u\in(-\beta,\beta)$
and by analytic continuation for other values of $t$ and $u$, where $\theta$ is the Heaviside step function. 
Depending on the spin structure chosen on the circle, the Euclidean propagator can be either periodic or antiperiodic in $x$ with period $L$ (these two alternative conditions are known respectively as Ramond and Neveu-Schwarz boundary conditions), and it is 
antiperiodic in $t$ with period $\beta$. Due to these quasiperiodicity properties, we may view $G$ as a section of a line bundle over a torus of circumferences $L$ and $\beta$. Taking the time derivative of the above equation and using (\ref{ham}) one obtains
\begin{equation}\label{diffeucl}
    (\partial_t\pm i\partial_x)G=\delta(x-y)\delta(t-u)
\end{equation}
for $(x,t)-(y,u)\in(-L,L)\times(-\beta,\beta)$. Identifying ${\mathbb R}^2$ with ${\mathbb C}$ via the map $(x,t)\mapsto x\mp it$, this equation says precisely that $G(z,w)$ is analytic in $z$ for 
$z\ne w$ and has a simple pole at $z=w$ with residue $\pm 1/(2\pi i)$ (the latter part of this statement is shown by integrating (\ref{diffeucl}) over a region containing $(y,u)$ and then applying Green's theorem). In other words,
\begin{equation}\label{eucleq}
    G(z,w)=\pm\frac{1}{2\pi i}\frac{1}{z-w}+F(z,w),
\end{equation}
where $F$ is analytic in $z$ for $z-w\in(-L,L)\times(-\beta,\beta)$. This equation is to be supplemented with the quasiperiodicity conditions
\begin{equation}\label{bc}
    G(z+P_i,w)=(-1)^{\nu_i}G(z,w)
\end{equation}
for $i=1,2$,
where $P_1=L$, $P_2=i\beta$, $\nu_1\in\{0,1\}$ and $\nu_2=1$. Eqs.~(\ref{eucleq}) and (\ref{bc}) have a unique solution. Indeed, the difference $\Delta G$ between two solutions, viewed as a function of its first argument,
is analytic for $z-w\in(-L,L)\times(-\beta,\beta)$ and satisfies (\ref{bc}), so it is analytic and bounded throughout the complex plane. By Liouville's theorem, such a function is necessarily a constant, so the antiperiodicity in the imaginary direction implies $\Delta G=0$. In order to find the solution, let us first look at the limiting case $L,\beta\to\infty$, where the torus becomes a plane and the quasiperiodicity conditions (\ref{bc}) are replaced by the condition that $G$ vanish at infinity. Since the only analytic function that vanishes at infinity is the zero function, the solution of Eq.~(\ref{eucleq}) on the plane is
\begin{equation}\label{propplane}
    G(z,w)=\pm\frac{1}{2\pi i}\frac{1}{z-w}\equiv G_0(z,w).
\end{equation}
Going back to the torus, i.e., to generic values of $L$ and $\beta$, we can solve Eqs.~(\ref{eucleq}) and (\ref{bc}) by the method of images,
\begin{equation}\label{images0}
    G(z,w)=\sum_{\lambda\in\Lambda}(-1)^{\nu\cdot\lambda}G_0(z+\lambda,w),
\end{equation}
where $\Lambda$ is the lattice
\begin{equation}
    \Lambda=\{\lambda_1 P_1+\lambda_2 P_2,\, \lambda_i\in{\mathbb Z}\}
\end{equation}
and $\nu\cdot\lambda=\nu_1\lambda_1+\nu_2\lambda_2=\nu_1\lambda_1+\lambda_2$.
Indeed, the function (\ref{images0}) clearly has the form (\ref{eucleq}), and one can easily check that it satisfies the quasiperiodicity conditions (\ref{bc}). This solution has to be taken as a formal one, because the above series is not absolutely convergent due to the slow decay of $G_0$. The argument can be made rigorous by adding a mass $m$, which makes the Euclidean propagator on the plane an exponentially decaying function, and letting $m\to 0$ at the end of the calculation.

\begin{figure}[t]
     \centering
     \includegraphics[scale=0.6]{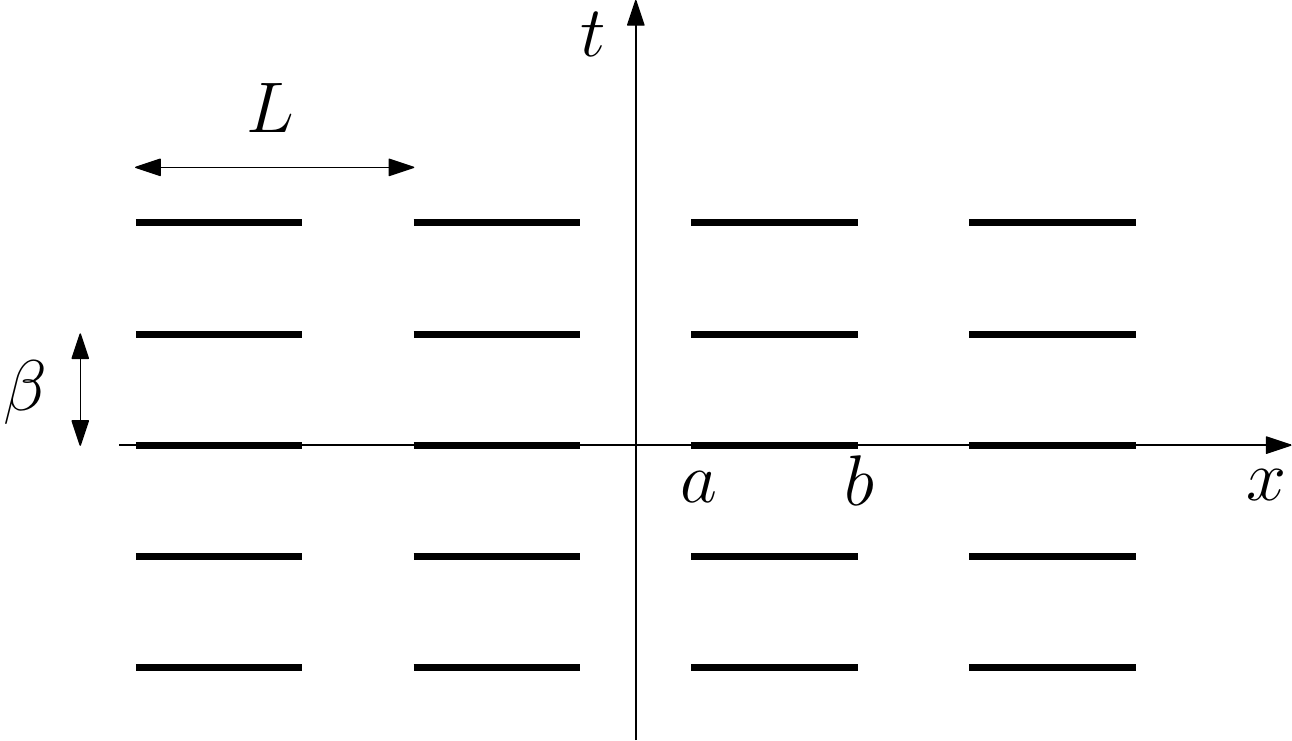}
     \caption{The region $V_\Lambda$ when $V$ is the interval $(a,b)$.}
     \label{fig1}
 \end{figure}

Let now $x,y\in V$. It is clear from (\ref{eucl}) that $G_V(x,y)=G(x,0^+;y,0)$ (note that it is important to take the limit $t\to 0$ from above, because if we take it from below we pick a delta function). With our identification of ${\mathbb R}^2$ with ${\mathbb C}$, we thus have
\begin{equation}\label{2pointprop}
    G_V(x,y)=G(x\mp i\epsilon,y),
\end{equation}
so, by (\ref{images0}),
\begin{equation}\label{images}
    G_V(x,y)=\sum_{\lambda\in{\Lambda}}(-1)^{\nu\cdot\lambda} G_{0V_\Lambda}(x+\lambda,y),
\end{equation}
where $V_\Lambda$ is a collection of segments distributed all over the complex plane,
\begin{equation}
    V_\Lambda=\bigcup_{\lambda\in\Lambda}(V+\lambda),
\end{equation}
and $G_{0V_\Lambda}$ is the function on $V_\Lambda\times V_\Lambda$ defined by
\begin{equation}
    G_{0V_\Lambda}(u,v)=G_0(u\mp i\epsilon,v).
\end{equation}
The region $V_\Lambda$ is represented in Fig.~\ref{fig1} in the case where $V$ is a single interval.
The main reason why the method of images is useful for us is that an equation analogous to (\ref{images}) holds for the powers of the operators involved,
\begin{equation}\label{imagesp}
     (G_{V}^n)(x,y)=\sum_{\lambda\in\Lambda}(-1)^{\nu\cdot\lambda}(G_{0V_\Lambda}^{n})(x+\lambda,y)
\end{equation}
for any $n\in{\mathbb N}$, where $(A^n)(u,v)$ denotes the kernel of the operator $A^n$ (not to be confused with $[A(u,v)]^n$). We can see this by induction. First, the above equation is satisfied for $n=1$ (this is Eq.~(\ref{images})). And second, if it holds for some $n\in{\mathbb N}$ we have
\begin{alignat}{2}
(G_{V}^{n+1})(x,y) &=\int_V dz\, G_{V}(x,z)\,(G_{V}^{n})(z,y)\nonumber\\
&=\sum_{\lambda,\mu\in\Lambda}(-1)^{\nu\cdot(\lambda+\mu)}\int_V dz\, G_{0V_\Lambda}(x+\lambda,z)\,( G_{0V_\Lambda}^n)(z+\mu,y)\nonumber\\
&= \sum_{\lambda,\mu\in\Lambda}(-1)^{\nu\cdot(\lambda+\mu)}\int_V dz\, G_{0V_\Lambda}(x+\lambda+\mu,z+\mu)\,(G_{0V_\Lambda}^n)(z+\mu,y)\nonumber\\
&= \sum_{\lambda',\mu\in\Lambda}(-1)^{\nu\cdot\lambda'}\int_{V+\mu}dz'\, G_{0V_\Lambda}(x+\lambda',z')\,( G_{0V_\Lambda}^{n})(z',y)\nonumber\\
&= \sum_{\lambda'\in\Lambda}(-1)^{\nu\cdot\lambda'}\int_{V_\Lambda}dz'\, G_{0V_\Lambda}(x+\lambda',z')\,( G_{0V_\Lambda}^n)(z',y)\nonumber\\
&=\sum_{\lambda'\in\Lambda}(-1)^{\nu\cdot\lambda'}(G_{0V_\Lambda}^{n+1})(x+\lambda',y),
\end{alignat}
which completes the proof. In the third equality we have used the translational invariance of $G_0$, and in the fourth we have defined $\lambda'=\lambda+\mu$ and $z'=z+\mu$. Eq.~(\ref{imagesp}) implies that the method of images works for any function of $G_V$ which can be expressed as a power series. In particular, it works for the resolvent,
\begin{equation}\label{resimages}
    R_{V}(\xi;x,y)=\sum_{\lambda\in\Lambda}(-1)^{\nu\cdot\lambda}R_{0V_\Lambda}(\xi;x+\lambda,y).
\end{equation}
In the case of zero temperature, $\beta\to\infty$, the terms with $\lambda_2\ne 0$ do not contribute to the sum (\ref{images0}), so the lattice $\Lambda$ effectively reduces to $\{mL,m\in{\mathbb Z}\}$ and, in consequence, the region $V_\Lambda$ reduces to an arrangement of segments in the real line. The resolvent $R_{0V_\Lambda}$ is well-known in that case \cite{Casini:2009vk}, so we can use it to obtain $R_V$ via the above equation. In \cite{Blanco:2019xwi} we showed that, in fact, $R_{0V_\Lambda}$ can be easily computed for generic temperatures, where $V_\Lambda$ is a collection of segments distributed all over the complex plane. In the next section we review the argument in a slightly simplified form.

\subsection{The resolvent for any set of segments in the plane}

\begin{figure}[t]
     \centering
     \includegraphics[scale=0.7]{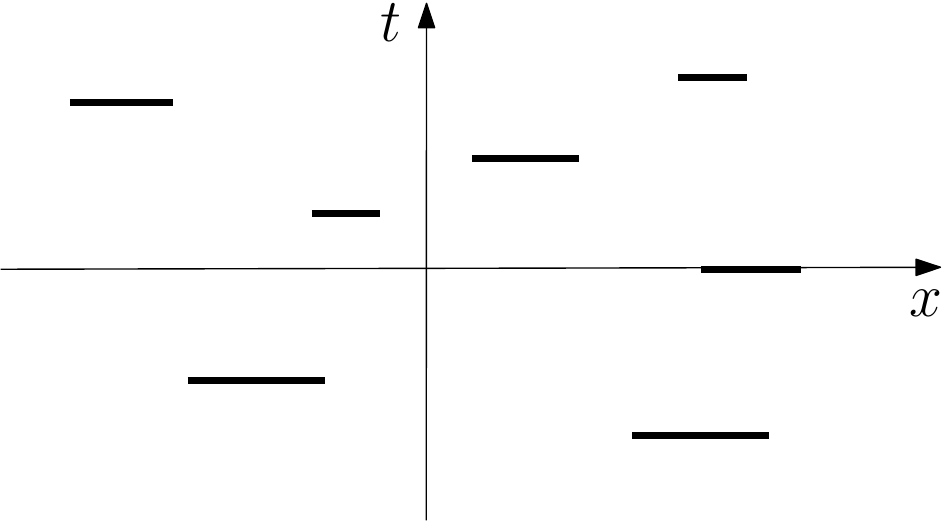}
     \caption{A collection of horizontal segments in the complex plane.}
     \label{fig2}
 \end{figure}

Let $A$ be a collection of horizontal segments in the complex plane, see Fig.~\ref{fig2}.
The purpose of this section is to compute the resolvent of the operator $G_{0A}$ with kernel
\begin{equation}\label{G0A}
    G_{0A}(u,v)=G_0(u\mp i\epsilon,v)=\pm\frac{1}{2\pi i}\frac{1}{u-v\mp i\epsilon}
\end{equation}
for $u,v\in A$. As we will see, the computation is very simple: we will first obtain an expression for the powers of $G_{0A}$, then we will insert it into the power expansion of the resolvent and find that it is easy to perform the sum. For the square we have
\begin{alignat}{2}\label{square}
    (G_{0A}^2)(u,v)&=\frac{1}{(2\pi i)^2}\int_A \frac{dw}{(u-w\mp i\epsilon)(w-v\mp i\epsilon)}\nonumber\\
    &=\frac{1}{(2\pi i)^2}\frac{1}{u-v\mp i\epsilon}\int_A dw\left(\frac{1}{u-w\mp i\epsilon}-\frac{1}{v-w\pm i\epsilon}\right)\nonumber\\
    &=\frac{1}{2\pi i}G_{0A}(u,v)\left[\omega_{A}^\pm(u)+\omega_{A}^\mp(v)\right],
\end{alignat}
where
\begin{equation}\label{omegaA}
    \omega_{A}^\pm(u)=\pm\int_A\frac{dw}{u-w\mp i\epsilon}.
\end{equation}
The second equality in (\ref{square}) is immediately checked by performing the difference of fractions. Of course, the above integral is easily computed (it gives a logarithm), but for future purposes it is simplest to momentarily keep the integral expression in order to avoid dealing with the branches of the complex logarithm. Note that
\begin{equation}\label{sumomega}
    \omega_{A}^++\omega_{A}^-=2\pi i.
\end{equation}
Indeed, if $A^{+/-}$ is a slight deformation of $A$ that circumvents the point $u\in A$ by taking the upper/lower path, we have
\begin{alignat}{2}
    \omega_{A}^+(u)+\omega_{A}^-(u)&=\int_A\frac{dw}{u-w- i\epsilon}-\int_A\frac{dw}{u-w+ i\epsilon}\nonumber\\
    &=\int_{A^+}\frac{dw}{u-w}-\int_{A^-}\frac{dw}{u-w}\nonumber\\
    &=\ointclockwise\frac{dw}{u-w}=2\pi i,
\end{alignat}
where the contour in the last integral encircles $u$. Eq.~(\ref{square}), which gives the kernel of $G_{0A}^2$, can be rewritten as an operator equation,
\begin{equation}\label{squareop}
    G_{0A}^2=\frac{1}{2\pi i}\left(\omega_{A}^\pm G_{0A}+G_{0A}\omega_{A}^\mp\right)
\end{equation}
Now, using (\ref{sumomega}) and (\ref{squareop}) it is a simple matter to check that the operator-valued function
\begin{equation}\label{barG}
    \bar G_{0A}(s)=(1+s)^{-\omega_{A}^\pm/(2\pi i)}\,G_{0A}\,(1+s)^{-\omega_{A}^\mp/(2\pi i)}
\end{equation}
satisfies $\bar G_{0A}'=-\bar G_{0A}^2$. In turn, this implies for the $n$-th derivative $\bar G_{0A}^{(n)}=(-1)^nn!\bar G_{0A}^{n+1}$, as can be easily shown by induction. Noting that $\bar G_{0A}(0)=G_{0A}$, we thus obtain
\begin{equation}
    G_{0A}^{n+1}=\frac{(-1)^n}{n!}\bar G_{0A}^{(n)}(0),
\end{equation}
which is the desired expression for the powers of $G_{0A}$. Inserting it into the power expansion of the resolvent (which is a geometric series), we recognize the Taylor series of $\bar G_{0A}$,
\begin{alignat}{2}
R_{0A}(\xi)&=\frac{1}{G_{0A}+\xi-1/2}\nonumber\\
 &=\frac{1}{\xi-1/2}\left[1-\frac{1}{\xi-1/2}\sum_{n=0}^\infty\frac{(-1)^n}{(\xi-1/2)^n}G_{0A}^{n+1}\right]\nonumber\\
 &=\frac{1}{\xi-1/2}\left[1-\frac{\bar G_{0A}(1/(\xi-1/2))}{\xi-1/2}\right]\nonumber\\
 &=\frac{1}{\xi-1/2}\left[1-\frac{e^{-ik(\xi)\omega_{A}^\pm}G_{0A}e^{ik(\xi)\omega_{A}^\pm}}{\xi+1/2}\right],
\end{alignat}
where in the last step we have used (\ref{sumomega}) to express $\omega_{A}^\mp$ in terms of $\omega_{A}^\pm$ and defined
\begin{equation}
    k(\xi)=\frac{1}{2\pi}\log\frac{\xi-1/2}{\xi+1/2}.
\end{equation}
Finally, using (\ref{G0A}) and the relation $1/(x\mp i\epsilon)=1/x\pm i\pi\delta(x)$ (there is a principal part implicit in the first term), we obtain
\begin{equation}\label{resol2}
    R_{0A}(\xi;u,v)=\frac{1}{\xi^2-1/4}\left\{\xi\delta(u-v)\mp \frac{e^{-ik(\xi)[\omega_{A}^\pm(u)-\omega_{A}^\pm(v)]}}{2\pi i(u-v)}\right\}.
\end{equation}
This is the resolvent for an arbitrary set $A$ of horizontal segments in the complex plane. It coincides with the result of \cite{Casini:2009vk} in the case where $A$ is contained in the real line.

\subsection{The resolvent on the torus}

Next we use the resolvent just computed, Eq.~(\ref{resol2}), to obtain the resolvent on the torus by the method of images, Eq.~(\ref{resimages}). From now on we will make extensive use of the Weierstrass functions, which are reviewed in appendix \ref{sec:weierstrass} (more extensive treatments are for example \cite{chandrasekharan2012elliptic,Pastras:2017wot}). We suggest the unfamiliar reader to go through it before continuing. Setting $A=V_\Lambda$ in (\ref{omegaA}) we obtain
\begin{alignat}{2}\label{omegalambda1}
    \omega_{V_\Lambda}^\pm(u)&=\pm\int_{V_\Lambda}\frac{dw}{u-w\mp i\epsilon}=\pm\sum_{\lambda\in\Lambda}\int_{V+\lambda}\frac{dw}{u-w\mp i\epsilon}\nonumber\\
    &=\pm\int_V dz\sum_{\lambda\in\Lambda}\frac{1}{u-z\mp i\epsilon-\lambda},
\end{alignat}
where in the last step we have defined $z=w-\lambda$. The series on the last member of this equation is ambiguous because it is not absolutely convergent, as can be easily seen by the integral test. However, its second derivative is unambiguous,
\begin{equation}
    (\omega_{V_\Lambda}^\pm)''(u)=\pm 2 \int_Vdz\sum_{\lambda\in\Lambda}\frac{1}{(u-z\mp i\epsilon-\lambda)^3}=\mp\int_V dz\,\wp'(u-z\mp i\epsilon),
\end{equation}
where $\wp$ is the Weierstrass elliptic function. Therefore,
\begin{equation}\label{omegapm}
    \omega_{V_\Lambda}^\pm(u)=\omega^\pm(u)\pm c u\pm d\qquad \omega^\pm(u)=\pm\int_V dz\,\zeta(u-z\mp i\epsilon)
\end{equation}
where $c$ and $d$ are undetermined constants (which parameterize the ambiguity we found in (\ref{omegalambda1})) and $\zeta$ is the Weierstrass zeta function. Note that property (\ref{sumomega}) and the fact that $\zeta$ has a simple pole with unit residue at the origin force $c$ and $d$ to be independent of the chirality.
Let now $N$ be the number of intervals in $V$, and let $a_i$ and $b_i$ be respectively the left and right endpoints of the $i$-th interval, that is,
\begin{equation}
    V=\bigcup_{i=1}^N(a_i,b_i).
\end{equation}
For $u=x\in V$, the integral in (\ref{omegapm}) is easily computed,
\begin{equation}\label{omega}
    \omega^\pm(x)=\pm\omega(x)+i\pi\qquad\omega(x)=\sum_{i=1}^N\log\left|\frac{\sigma(a_i-x)}{\sigma(b_i-x)}\right|,
\end{equation}
where $\sigma$ is the Weierstrass sigma function (the $i\pi$ term comes from the interval containing $x$, which passes near a pole of the integrand). On the other hand, the quasiperiodicity of $\zeta$, Eq.~(\ref{zetap}), implies
\begin{equation}\label{omegap}
    \omega^\pm(x+\lambda)=\omega^\pm(x)\pm 2 \ell\lambda\cdot\zeta(P/2),
\end{equation}
where $\ell=\int_V dz$ is the total length of $V$ and $\lambda\cdot\zeta(P/2)=\lambda_1\zeta(P_1/2)+\lambda_2\zeta(P_2/2)$. Setting $A=V_\Lambda$ in the resolvent (\ref{resol2}), substituting into (\ref{resimages}) and using (\ref{omegapm}), (\ref{omega}) and (\ref{omegap}), we obtain for the resolvent on the torus
\begin{equation}\label{resoltorus}
    R_V(\xi;x,y)=\frac{1}{\xi^2-1/4}\left\{\xi\delta(x-y)-e^{\mp ik(\xi)[\omega(x)-\omega(y)]}F^\pm(\xi;x,y)\right\},
\end{equation}
where
\begin{equation}\label{Fpm}
    F^\pm(\xi;z,w)=\pm\frac{1}{2\pi i}\sum_{\lambda\in\Lambda}(-1)^{\nu\cdot\lambda}\frac{e^{\mp i k(\xi)[2\ell\lambda\cdot\zeta(P/2)+c(z-w+\lambda)]}}{z-w+\lambda}.
\end{equation}
Note that the constant $d$ has dropped out (this is because $R_{0V_\Lambda}$ only involves the difference $\omega_{V_\Lambda}^\pm(u)-\omega_{V_\Lambda}^\pm(v)$, see (\ref{resol2})), but the constant $c$ remains. The latter is fixed by requiring the summand above not to diverge exponentially as $|\lambda|$ grows. Indeed, taking into account that $\zeta$ is odd and satisfies $\zeta(z^*)=\zeta^*(z)$, so that $\zeta(L/2)$ is real and $\zeta(i\beta/2)$ is imaginary, one easily sees that the only way to prevent the exponential divergence is to set
\begin{equation}
    c=-\frac{2\ell}{i\beta}\zeta(i\beta/2),
\end{equation}
so that the ambiguity in $\omega^\pm_{V_\Lambda}$ disappears from the resolvent on the torus. 
With this choice of $c$, the summand in (\ref{Fpm}) decays to zero as $|\lambda|$ grows, but not quickly enough to make the series absolutely convergent, so there remains an ambiguity. However, note that, formally, $F^\pm$ has the following properties: (i) it is analytic in $z$ for $z-w\in(-L,L)\times(-\beta,\beta)$ except for a simple pole with residue $\pm 1/(2\pi i)$ at $z=w$; and (ii) it is quasiperiodic,
\begin{equation}
    F(\xi;z+P_i,w)=(-1)^{\nu_i}e^{\pm 2ik(\xi)\ell\zeta(P_i/2)}F^\pm(\xi;z,w).
\end{equation}
By the argument we gave under Eq.~(\ref{bc}), there is only one function with these properties, so imposing them cures the ambiguity. The result is
\begin{equation}\label{F}
    F^\pm(\xi;z,w)=\pm\frac{1}{2\pi i}\frac{1}{\sigma(z-w)}\frac{\sigma_\nu(z-w\pm ik(\xi)\ell)}{\sigma_\nu(\pm ik(\xi)\ell)},
\end{equation}
where
\begin{equation}
    \sigma_\nu(z)=e^{-[\nu_1\zeta(P_2/2)+\nu_2\zeta(P_1/2)]z}\sigma(z+\nu_1P_2/2+\nu_2P_1/2)
\end{equation}
(recall that $\nu_2=1$). Indeed, one can easily check from the properties of $\sigma$ discussed in appendix \ref{sec:weierstrass} that the function (\ref{F}) satisfies conditions (i) and (ii) above. To verify the latter condition one has to use the quasiperiodicity property
\begin{equation}\label{sigmanup}
    \sigma_\nu(z+P_i)=(-1)^{\nu_i+1}e^{\zeta(P_i/2)(2z+P_i)}\sigma_\nu(z),
\end{equation}
which follows from the quasiperiodicity property of $\sigma$, Eq.~(\ref{sigmap}), and the relation (\ref{halfperiods}) between the values of $\zeta$ at the half-periods. From (\ref{resoltorus}) and (\ref{F}) we finally obtain
\begin{equation}\label{resoltorusfinal}
    R_V(\xi;x,y)=\frac{1}{\xi^2-1/4}\left\{\xi\delta(x-y)\mp\frac{e^{\mp ik(\xi)[\omega(x)-\omega(y)]}}{2\pi i\sigma(x-y)}\frac{\sigma_\nu(x-y\pm ik(\xi)\ell)}{\sigma_\nu(\pm ik(\xi)\ell)}\right\},
\end{equation}
which is the resolvent on the torus and thus the main result of this section.

\section{Modular Hamiltonian}\label{sect:3}

\begin{figure}[t]
    \centering
           \includegraphics[scale=0.7]{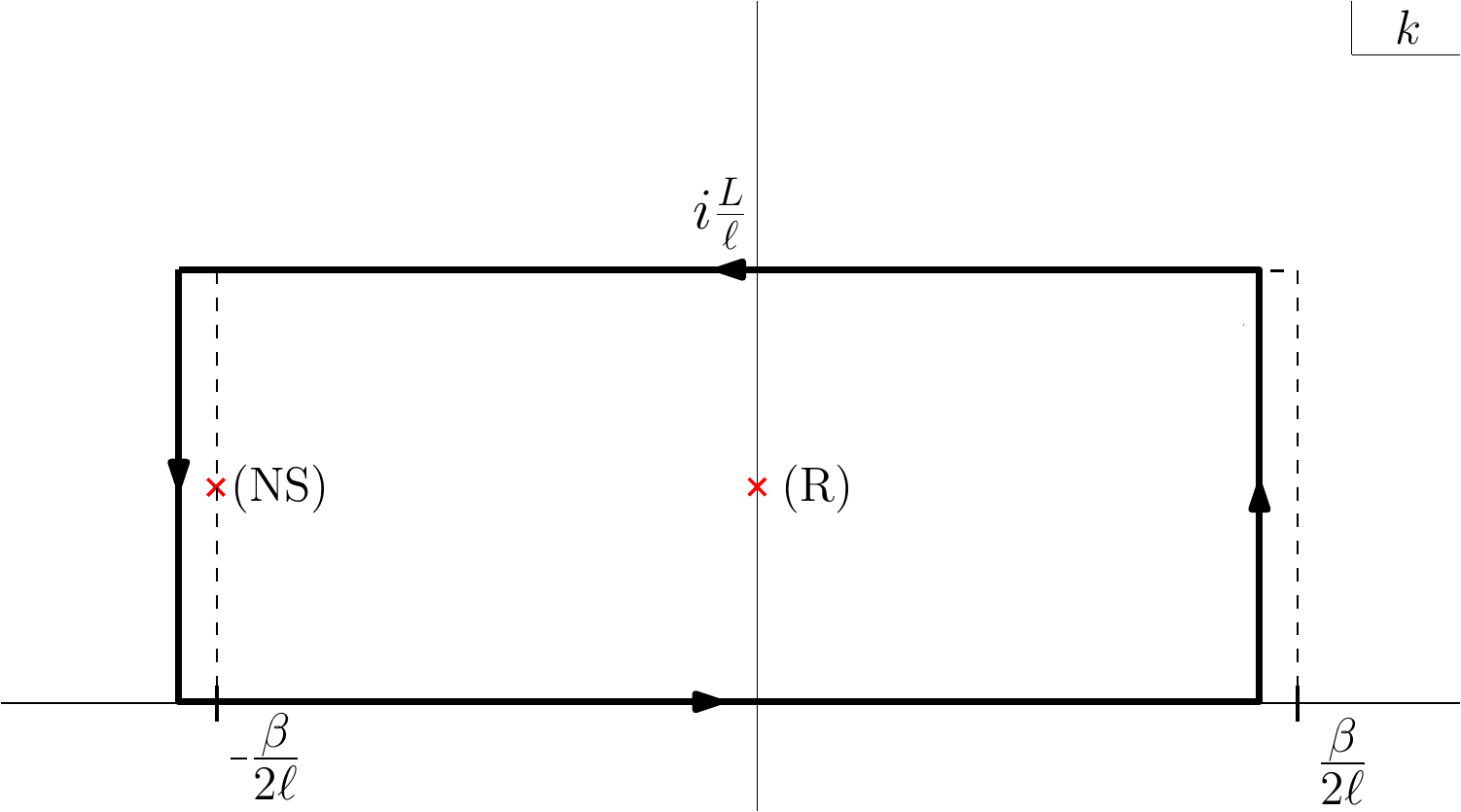} 
        \caption{Integration contour of Eq.~(\ref{contourint}). Also displayed is the pole of the integrand, which depends on the boundary conditions.}  \label{fig:circuito}
\end{figure}

\begin{figure}[t]
    \centering
           \includegraphics[width=\linewidth]{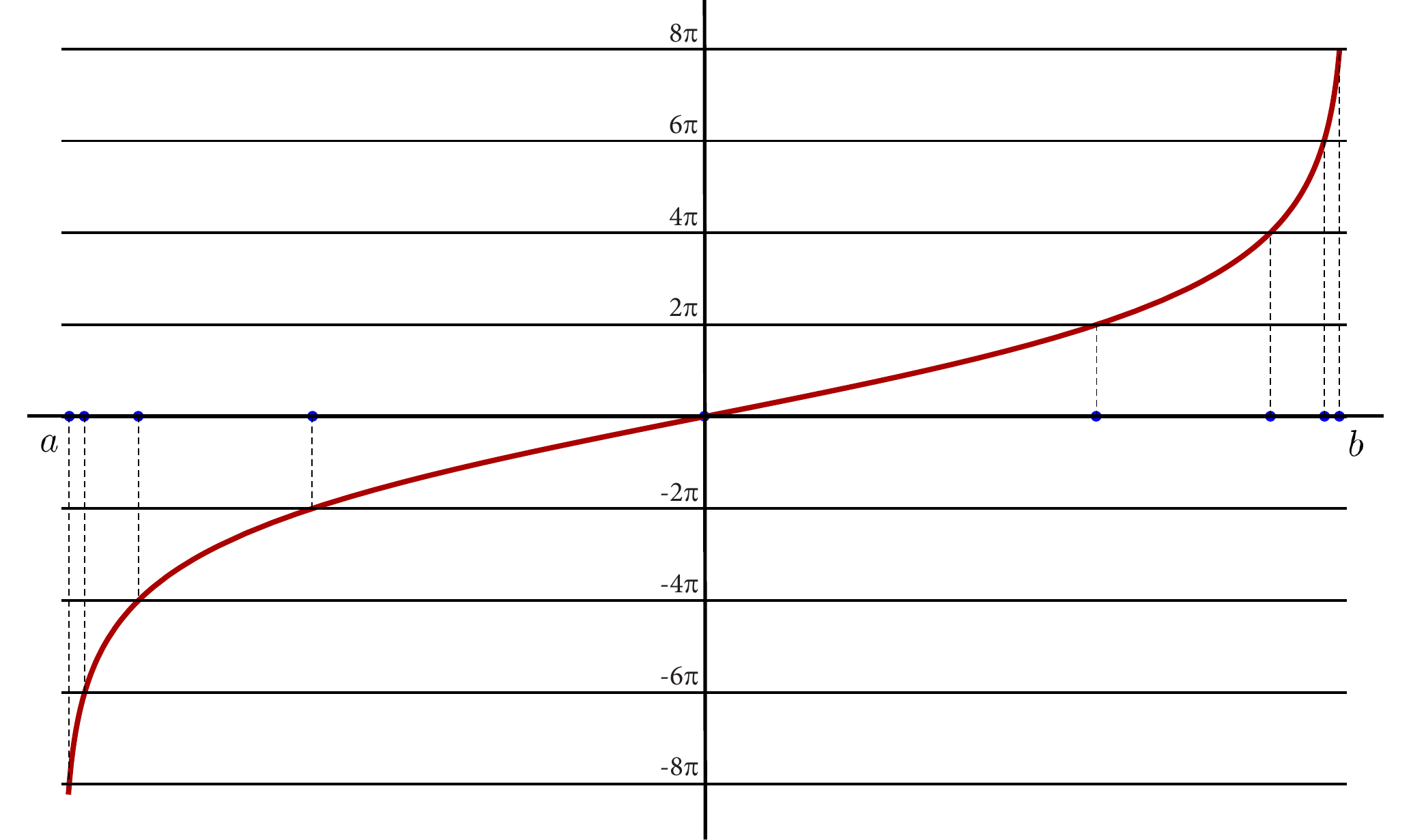} 
        \caption{The function $g(y)=\beta[z(y)-z(x)]/\ell$, with $z$ defined in (\ref{z}), in the case where $V$ is the interval $(a,b)$ and $x$ is the midpoint of the interval. Also displayed are the first few points $x_n$ satisfying $g(x_n)=2\pi n$, which the modular Hamiltonian (\ref{moddec})-(\ref{nonlocal}) couples to $x$. They tend to accumulate near the endpoints of the interval.} \label{fig:ceros}
\end{figure}

We can now obtain the modular Hamiltonian by substituting the resolvent we just computed, Eq.~(\ref{resoltorusfinal}), into (\ref{modres}). Note that the term with a delta function in (\ref{resoltorusfinal}) cancels from the modular Hamiltonian, because it changes sign under the transformation $\xi\mapsto -\xi$. Changing the variable of integration from $\xi$ to $k(\xi)$ we find
\begin{alignat}{2}\label{modtorus}
    &K_V(x,y)=\frac{\mp i}{\sigma(x-y)}\int_{-\infty}^\infty dk\, f(k;x,y)\nonumber\\ &f(k;x,y)=e^{-ik[\omega(x)-\omega(y)]}\frac{\sigma_\nu(x-y+ik\ell)}{\sigma_\nu(ik\ell)}.
\end{alignat}
Surprisingly, the integral above is easily computed. To see this, note first from (\ref{sigmanup}) that $f$ is quasiperiodic in $k$,
\begin{alignat}{2}\label{fp}
    &f(k+\beta/\ell;x,y)=e^{-i\frac{\beta}{\ell}[z(x)-z(y)]}f(k;x,y)\nonumber\\
    &f(k+iL/\ell;x,y)=e^{\frac{L}{\ell}[z(x)-z(y)]-\frac{2\pi}{\beta}(x-y)}f(k;x,y),
\end{alignat}
where
\begin{equation}\label{z}
    z(x)=\omega(x)-\frac{2\ell}{i\beta}\zeta(i\beta/2)x=\sum_{i=1}^N\log\left|\frac{\sigma(a_i-x)}{\sigma(b_i-x)}\right|-\frac{2\ell}{i\beta}\zeta(i\beta/2)x
\end{equation}
(in deriving the second equation in (\ref{fp}) we have also used the relation (\ref{halfperiods}) between the values of $\zeta$ at the half-periods). Note that $z$ is real because $\zeta(i\beta/2)$ is imaginary. Note also that $z(x)\to \pm\infty$ as $x$ approaches the right/left endpoint of any interval of $V$, because $\sigma$ vanishes at the origin. In appendix \ref{sec:monotonicity} we show that, in fact, $z$ is a monotonically increasing function on each interval. Using the first of the above quasiperiodicity properties, we can rewrite the integral in (\ref{modtorus}) as
\begin{alignat}{2}\label{integralp}
       \int_{-\infty}^\infty dk\, f(k;x,y)&=\sum_{n\in{\mathbb Z}}\int_{-\frac{\beta}{2\ell}}^{\frac{\beta}{2\ell}}dk\,f(k+n\beta/\ell;x,y)\nonumber\\
     &=\int_{-\frac{\beta}{2\ell}}^{\frac{\beta}{2\ell}} dk\, f(k;x,y)\sum_{n\in{\mathbb Z}}e^{-in\frac{\beta}{\ell}[z(x)-z(y)]}\nonumber\\
     &=2\pi\int_{-\frac{\beta}{2\ell}}^{\frac{\beta}{2\ell}} dk\, f(k;x,y)\sum_{n\in{\mathbb Z}}\delta\left(\frac{\beta}{\ell}[z(x)-z(y)]+2\pi n\right).
\end{alignat}
Except for a factor $2\pi$, the sum in the second line is the Fourier series of the delta function on the circle, hence the last equality.
Suppose now that $x$ is fixed in the $i$-th interval, $x\in(a_i,b_i)$. Since $z$ grows monotonically from $-\infty$ to $\infty$ on each interval,
for each $n\in{\mathbb Z}$ and $j\in\{1,\dots,N\}$ the argument of the delta function has a unique zero $y=x_{nj}$ lying in the $j$-th interval. In other words, there is a unique $x_{nj}\in(a_j,b_j)$ such that
\begin{equation}\label{rootsdelta}
    \frac{\beta}{\ell}[z(x)-z(x_{nj})]+2\pi n=0.
\end{equation}
Clearly, $x_{0i}=x$. Noting from (\ref{modtorus}) that $f(k;x,x)=1$, we may thus write
\begin{alignat}{2}\label{integral}
     \int_{-\infty}^\infty dk\, f(k;x,y)&=\frac{2\pi \ell}{\beta}\sum_{n\in{\mathbb Z}}\sum_{j=1}^N\frac{\delta(y-x_{nj})}{z'(x_{nj})}\int_{-\frac{\beta}{2\ell}}^{\frac{\beta}{2\ell}}dk\,f(k;x,x_{nj})\nonumber\\
     &=2\pi\frac{\delta(y-x)}{z'(x)}+\frac{2\pi \ell}{\beta}\sum_{(n,j)\ne(0,i)}\frac{\delta(y-x_{nj})}{z'(x_{nj})}\int_{-\frac{\beta}{2\ell}}^{\frac{\beta}{2\ell}}dk\,f(k;x,x_{nj}).
\end{alignat}
Let us compute the remaining integral. The quasiperiodicity properties (\ref{fp}), together with (\ref{rootsdelta}), imply
\begin{alignat}{2}\label{fproots}
&f(k+\beta/\ell;x,x_{nj})=f(k;x,x_{nj})\nonumber\\
    &f(k+iL/\ell;x,x_{nj})=e^{-\frac{2\pi}{\beta}(x-x_{nj}+nL)}f(k;x,x_{nj})
\end{alignat}
and therefore, for the contour depicted in Fig.~\ref{fig:circuito}, we have
\begin{equation}\label{contourint}
    \ointctrclockwise dk\,f(k;x,x_{nj})=\left[1-e^{-\frac{2\pi}{\beta}(x-x_{nj}+nL)}\right]\int_{-\frac{\beta}{2\ell}}^{\frac{\beta}{2\ell}}dk\,f(k;x,x_{nj})
\end{equation}
(the contributions from the two vertical sides of the contour cancel because of the first equation in (\ref{fproots})).
The integrand has only one pole inside the contour, which comes from a zero of the denominator in (\ref{modtorus}) and is located at $k=-\nu_1\beta/(2\ell)+iL/(2\ell)$, as shown in the figure. Computing the contour integral by residues yields
\begin{equation}
    \int_{-\frac{\beta}{2\ell}}^{\frac{\beta}{2\ell}}dk\,f(k;x,x_{nj})=\frac{\pi}{\ell}\frac{(-1)^{n\nu_1}\sigma(x-x_{nj})}{\sinh\left[\frac{\pi}{\beta}(x-x_{nj}+nL)\right]},
\end{equation}
and inserting this result into (\ref{integral}) we obtain
\begin{equation}\label{integral2}
    \int_{-\infty}^\infty dk\,f(k;x,y)=2\pi\frac{\delta(y-x)}{z'(x)}
    +\frac{2\pi^2}{\beta}\sum_{(n,j)\ne(0,i)}\frac{(-1)^{n\nu_1}\sigma(x-x_{nj})\delta(y-x_{nj})}{z'(x_{nj})\sinh\left[\frac{\pi}{\beta}(x-x_{nj}+nL)\right]}.
\end{equation}
We can now substitute this expression into (\ref{modtorus}) to obtain the modular Hamiltonian. We find
that the modular Hamiltonian is the sum of a local term and a non-local term,
\begin{equation}\label{moddec}
    K_V=K_{V}^{(l)}+K_{V}^{(nl)},
\end{equation}
which come respectively from the first and second terms in (\ref{integral2}). The local term is proportional to the product of distributions $\delta(y-x)/(x-y)$ and needs to be regularized. As explained in \cite{Casini:2009vk}, the most general regularization gives $\delta(y-x)/(x-y)=\delta'(y-x)+k(x)\delta(y-x)$, and the arbitrary function $k$ can be fixed by requiring $K_V$ to be Hermitian. The result is
\begin{alignat}{2}
&K_{V}^{(l)}(x,y)=\pm i\left[-\tilde\beta(x)\delta'(y-x)+\frac{1}{2}\tilde\beta'(x)\delta(y-x)\right]\label{local}\\
&K_{V}^{(nl)}(x,y)=\mp \frac{i\pi}{\beta}\sum_{(n,j)\ne(0,i)}\frac{(-1)^{n\nu_1}\tilde\beta(x_{nj})}{\sinh\left[\frac{\pi}{\beta}(x-x_{nj}+nL)\right]}\delta(y-x_{nj})\label{nonlocal}
\end{alignat}
for $x\in(a_i,b_i)$, where we have defined
\begin{equation}\label{localtemp}
    \tilde\beta(x)=\frac{2\pi}{z'(x)}.
\end{equation}
Eqs.~(\ref{moddec})-(\ref{nonlocal}) give the modular Hamiltonian of a chiral fermion on the circle at non-zero temperature, for an arbitrary collection of intervals.
Comparing the local term with the dynamical Hamiltonian (\ref{ham}), it is clear that $\tilde\beta$ plays the role of a local inverse temperature.
Note that the non-local term survives even in the case of a single interval, where the condition $(n,j)\ne(0,i)$ reduces to $n\ne 0$. In that case, for $x$ fixed, it has support in an infinite number of points, which tend to accumulate near the endpoints of the interval as shown in Fig.~\ref{fig:ceros}.

\subsection{Limits}

Let us now study some limits of the above result, to understand it better and to compare with known results in the literature. Consider first the limit $L\to\infty$, with $\beta$ finite. From Eqs.~(\ref{sigmalim}) and (\ref{zetalim}) we see that, in this limit, (\ref{z}) becomes
\begin{equation}
    z(x)=\sum_{i=1}^N\log\left|\frac{\sinh\left[\frac{\pi}{\beta}(a_i-x)\right]}{\sinh\left[\frac{\pi}{\beta}(b_i-x)\right]}\right|,
\end{equation}
where we have dropped an irrelevant additive constant. On the other hand, in this limit the terms with $n\ne 0$ do not contribute to the sum in (\ref{nonlocal}), so the non-local term takes the form
\begin{equation}
    K_{V}^{(nl)}(x,y)=\mp \frac{i\pi}{\beta}\sum_{j\ne i}\frac{\tilde\beta(x_{j})}{\sinh\left[\frac{\pi}{\beta}(x-x_{j})\right]}\delta(y-x_{j}),
\end{equation}
where we have used the notation $x_j=x_{0j}$. This gives the modular Hamiltonian of a chiral fermion on the line at non-zero temperature, for an arbitrary number of intervals. Note that the dependence on $\nu_1$, i.e., on the boundary conditions, drops out in this limit, as it should be. Further taking the limit $\beta\to\infty$ of these expressions, one recovers the result of \cite{Casini:2009vk} for the modular Hamiltonian of chiral fermions on the line at zero temperature. 

The limit $\beta\to\infty$, with $L$ finite, is more subtle. From Eqs.~(\ref{sigmalim}) and (\ref{zetalim}) we see that, in this limit, (\ref{z}) takes the form
\begin{equation}
    z(x)=\sum_{i=1}^N\log\left|\frac{\sin\left[\frac{\pi}{L}(a_i-x)\right]}{\sin\left[\frac{\pi}{L}(b_i-x)\right]}\right|,
\end{equation}
where again we have dropped an irrelevant additive constant. In order to properly take the limit of the non-local term, it is convenient to first apply it to a function $f$ on $V$,
\begin{equation}\label{nonlocalf}
    (K_{V}^{(nl)}f)(x)=\mp \frac{i\pi}{\beta}\sum_{(n,j)\ne(0,i)}\frac{(-1)^{n\nu_1}\tilde\beta(x_{nj})}{\sinh\left[\frac{\pi}{\beta}(x-x_{nj}+nL)\right]}f(x_{nj}).
\end{equation}
Let us first compute the limit $\beta\to\infty$ of the partial sum $K_{V,1}^{(nl)}f$ of terms with $|n|L\ll\beta$.
Noting from (\ref{rootsdelta}) that, for these terms, $x_{nj}=x_{0j}\equiv x_j$, we find
\begin{equation}
     (K_{V,1}^{(nl)}f)(x)=\mp i\left[\tilde\beta(x_i)f(x_i)\sum_{n\ne 0}\frac{(-1)^{n\nu_1}}{nL}+\sum_{j\ne i}\tilde\beta(x_j)f(x_j)\sum_{n\in{\mathbb Z}}\frac{(-1)^{n\nu_1}}{x-x_j+nL}\right],
\end{equation}
where the infinite sums, which are not absolutely convergent, have to be taken in symmetric order because $|n|$ is bounded by $\beta$. The first series above vanishes, and the second gives a trigonometric function that depends on $\nu_1$. The result is
\begin{alignat}{2}
    K_{V,1}^{(nl)}(x,y)=\mp\frac{i\pi}{L}\sum_{j\ne i}\tilde\beta(x_j)g_{\nu_1}(x-x_j)\delta(y-x_j)
\end{alignat}
with
\begin{equation}
    g_0(u)=\cot(\pi u/L)\qquad g_1(u)=\frac{1}{\sin(\pi u/L)},
\end{equation}
in agreement with \cite{Wong:2013gua}. What remains to be computed is the contribution $K_{V,2}^{(nl)}f$ to (\ref{nonlocalf}) from terms with $|n|L\sim\beta$ (the terms with $|n|L\gg \beta$ do not contribute in the limit $\beta\to\infty$). Note first that, in this regime, the first two terms of the argument of the $\sinh$ in (\ref{nonlocalf}) can be dropped, because $x-x_{nj}$ is bounded by $L$ and $\beta$ is arbitrarily large. The remaining term can be rewritten using (\ref{rootsdelta}) as
\begin{equation}
    \frac{\pi nL}{\beta}=\frac{L}{2\ell}\left[z(x_{nj})-z(x)\right].
\end{equation}
Note also, using (\ref{rootsdelta}) again, that $z(x_{(n+1)j})-z(x_{nj})=2\pi \ell/\beta$. In the limit $\beta\to\infty$, this equation says that $x_{(n+1)j}$ is very close to $x_{nj}$, so we can write
\begin{equation}
    \Delta x_{nj}\equiv x_{(n+1)j}-x_{nj}=\frac{2\pi}{z'(x_{nj})}\frac{\ell}{\beta}=\tilde\beta(x_{nj})\frac{\ell}{\beta}.
\end{equation}
Graphically, what happens for $\beta$ large is that the curve plotted in Fig.~\ref{fig:ceros} becomes arbitrarily steep, so the different solutions of (\ref{rootsdelta}) approach each other.
Using these equations in (\ref{nonlocalf}) we obtain
\begin{equation}
    (K_{V,2}^{(nl)}f)(x)=\mp \frac{i\pi}{\ell}\sum_{(n,j)\ne(0,i)}\frac{(-1)^{n\nu_1}\Delta x_{nj}}{\sinh\left[\frac{L}{2\ell}(z(x_{nj})-z(x))\right]}f(x_{nj}).
\end{equation}
Now, since $\Delta x_{nj}$ is vanishingly small in the limit $\beta\to\infty$, in the Neveu-Schwarz case $\nu_1=1$ this sum vanishes because consecutive terms cancel each other. Contrarily, in the Ramond case $\nu_1=0$ the above sum becomes an integral, so we finally obtain
\begin{equation}
    K_{V,2}^{(nl)}(x,y)=\mp \frac{i\pi}{\ell}\frac{1}{\sinh\left[\frac{L}{2\ell}(z(y)-z(x))\right]},
\end{equation}
in agreement with \cite{Wong:2013gua}. This emergence of continuous non-locality had already been noticed for the case of a single interval in \cite{Fries:2019ozf}.

\begin{figure}[t]
    \centering
           \includegraphics[width=0.8\linewidth]{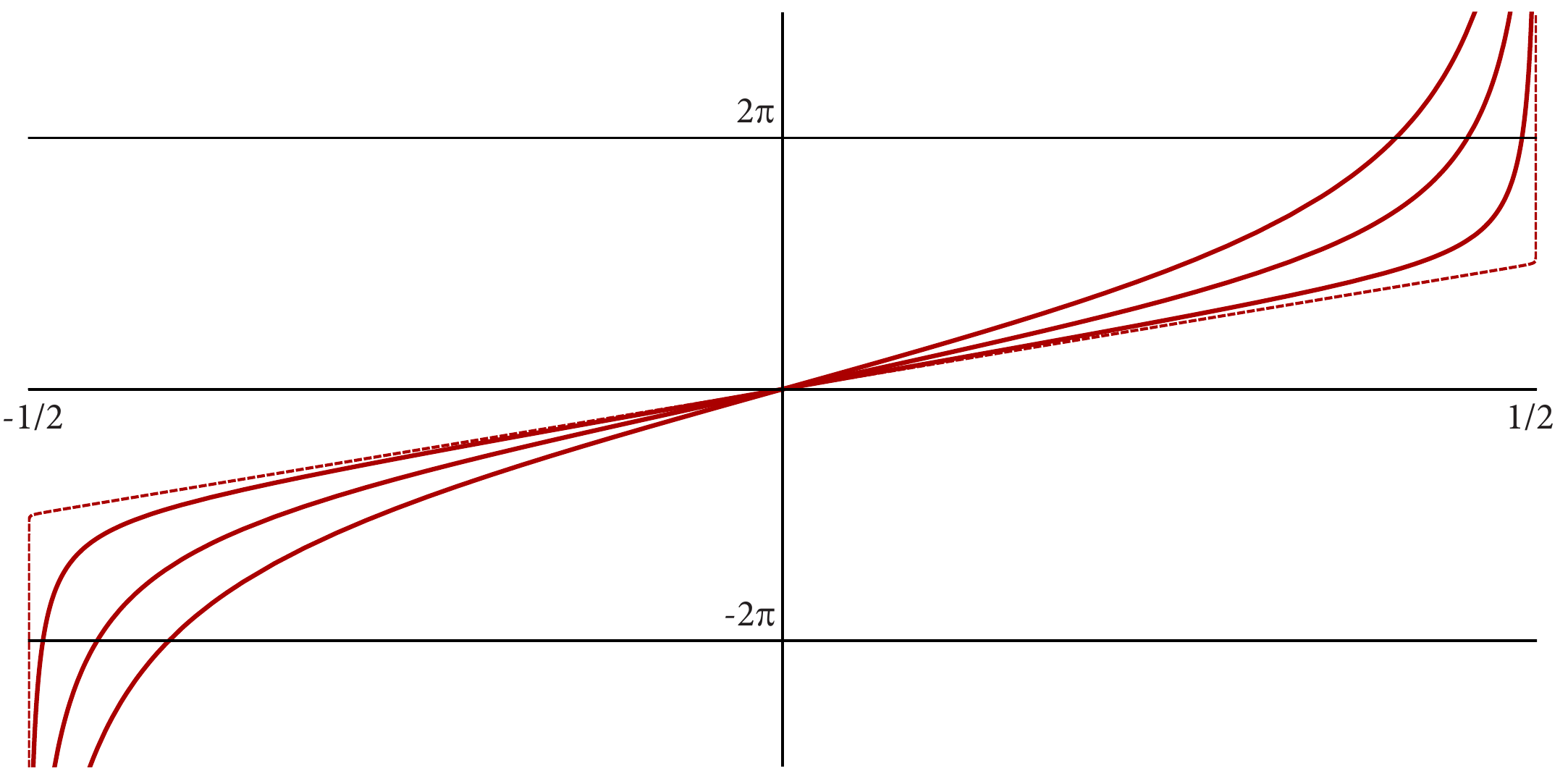} 
        \caption{The function $g(y)=\beta[z(y)-z(x)]/\ell$ in the case $V=(-\ell/2,\ell/2)$ and $x=0$, plotted against $y/\ell$ for different values of $\ell$. The dotted line corresponds to the limit $\ell\to L$. As $\ell$ increases, the solutions of the equation $g(y)=2\pi n$ approach the endpoints of the interval.} \label{fig:ceros2}
\end{figure}

Let us finally consider the case of a single interval, $V=(a,b)$, and study the limit $\ell\to L$, i.e., the limit in which $V$ becomes the whole circle, for arbitrary values of $L$ and $\beta$. From the quasiperiodicity property (\ref{sigmap}) of $\sigma$ and the relation (\ref{halfperiods}) between the values of $\zeta$ at the half-periods, we see that, in this limit, (\ref{z}) reduces to
\begin{equation}
    z(x)=\frac{2\pi x}{\beta}
\end{equation}
(where once again we have dropped an irrelevant additive constant),
so the local temperature equals the true temperature, $\tilde\beta=\beta$. Note that, in this limit, $z(x)$ does not diverge as $x$ approaches the endpoints of the interval: the zeros of $\sigma(a-x)$ and $\sigma(b-x)$ cancel each other. In fact, the values of $z$ at the endpoints are so small that Eq.~(\ref{rootsdelta}) has no solutions inside the interval. What happens, as shown in Fig.~\ref{fig:ceros2}, is that, as $\ell$ approaches $L$, the solutions of that equation approach the boundary of the interval, so that, in the limit, they have all reached the boundary. For $x\to a^-,b^+$ we have $z'(x)\to\infty$ and hence $\tilde\beta(x)\to 0$. It follows that the non-local term (\ref{nonlocal}) vanishes in the limit $\ell\to L$, so the modular Hamiltonian becomes the dynamical Hamiltonian times $\beta$, as it should be.

\subsection{Modular flow}

\begin{figure}[t]
    \centering
           \includegraphics[width=\linewidth]{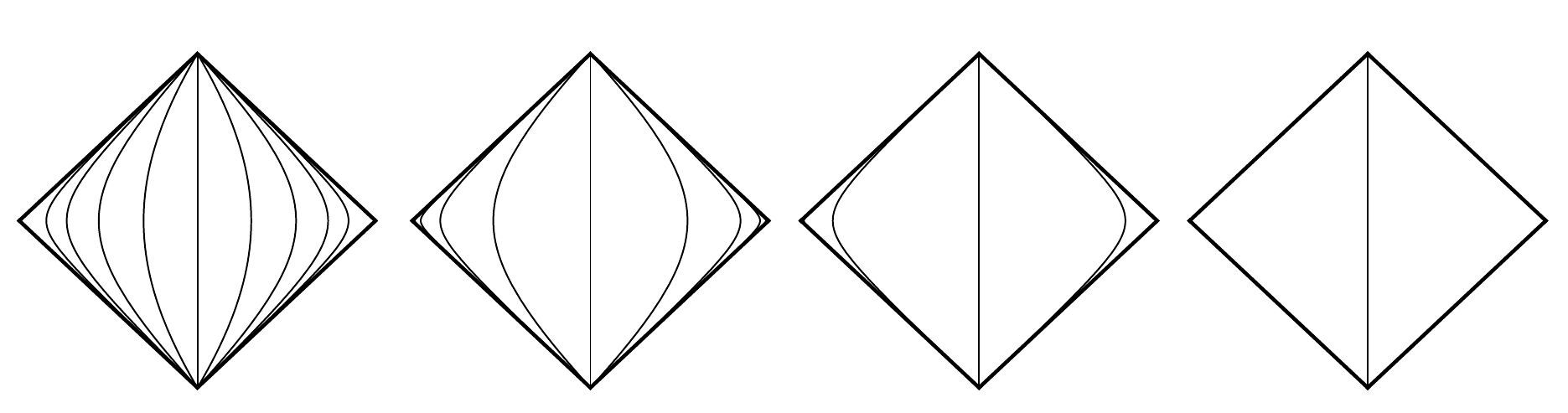} 
        \caption{The curves (\ref{curves}) describing the modular evolution of $\psi_D(x)$, in the case where $x$ is the midpoint of the interval, for $L=10$, $\beta=10$ and $\ell=1,\,2,\,8,\,9$ from left to right. As $\ell\to L$, the non-locality tends to disappear and the modular evolution becomes time evolution, consistently with the fact that, in this limit, the modular Hamiltonian reduces to the dynamical Hamiltonian times $\beta$.} \label{fig:diamantes}
\end{figure}

\begin{figure}[t]
    \centering
           \includegraphics[width=\linewidth]{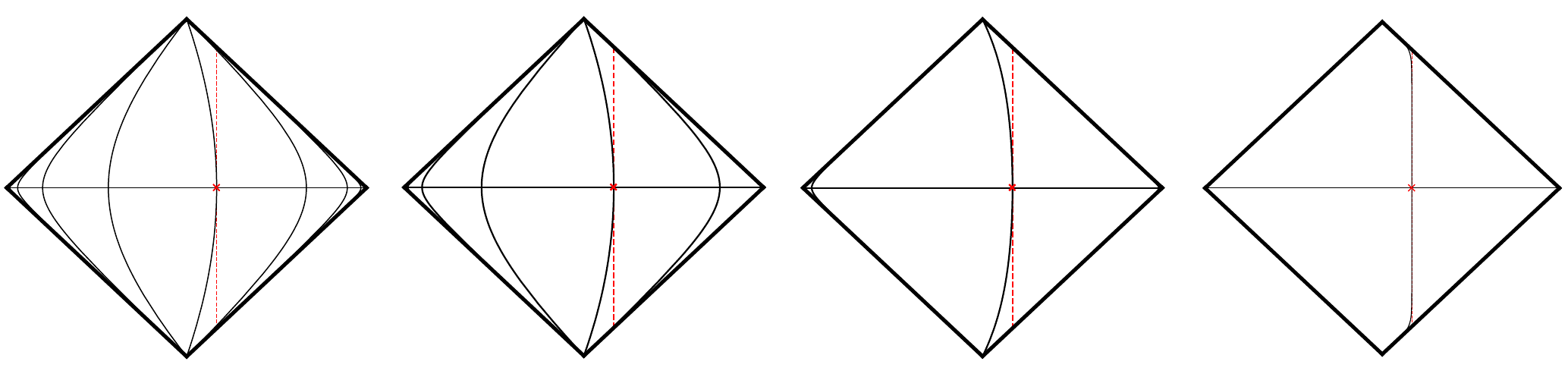} 
        \caption{The curves describing the modular evolution of $\psi_D(x)$ with $x$ displaced $\ell/12$ with respect to the midpoint of the interval, for $L=10$, $\ell=4$ 
        and $\beta=20,\,15,\,5,\,1/2$ from left to right. As the temperature increases, thermal effects start dominating over entanglement and, accordingly, modular evolution approaches time evolution.} \label{fig:diamantes2}
\end{figure}

Let us compute the modular evolution $\Psi_V$ of $\psi$, which is defined as
\begin{equation}
    \Psi_V(x,s)=e^{iH_Vs}\psi(x)e^{-iH_Vs},
\end{equation}
i.e., analogously to the Heisenberg evolution with $H_V$ replacing $H$. For simplicity, we will restrict to the case where $V$ is a single interval. 
The case of several intervals can be treated similarly.
Differentiating the above equation with respect to $s$ and using (\ref{moddec})-(\ref{nonlocal}), we obtain
\begin{alignat}{2}\label{flow1}
    \partial_s\Psi_V&=i[H_V,\Psi_V]=-iK_V\Psi_V\nonumber\\
    &=\pm\left\{\tilde\beta\partial_x\Psi_V+\frac{1}{2}\tilde\beta'\Psi_V-\frac{\pi}{\beta}\sum_{n\ne 0}\frac{(-1)^{n\nu_1}\tilde\beta(x_{n})}{\sinh\left[\frac{\pi}{\beta}(x-x_{n}+nL)\right]}\Psi_V(x_{n})\right\},
\end{alignat}
where we have dropped the subscript $j$ from $x_{nj}$ (the solution of Eq.~(\ref{rootsdelta}) in the $j$-th interval) because there is only one interval. With this notation, recall that $x_0=x$.
The above is a non-local equation, but it can be rewritten as a system of coupled local equations by defining $\Psi_m(x,s)\equiv\Psi_V(x_m(x),s)$.
Noting from (\ref{rootsdelta}) and (\ref{localtemp}) that $x_{m}'(x)=z'(x)/z'(x_m)=\tilde\beta(x_m)/\tilde\beta(x)$, we find
\begin{equation}\label{flow2}
    \partial_s\Psi_m=\pm\left\{\tilde\beta\partial_x\Psi_m+\frac{1}{2}\tilde\beta'(x_m)\Psi_m-\frac{\pi}{\beta}\sum_{n\ne m}\frac{(-1)^{(m-n)\nu_1}\tilde\beta(x_{n})}{\sinh\left[\frac{\pi}{\beta}(x_m-x_{n}-(m-n)L)\right]}\Psi_n\right\},
\end{equation}
which, after the change of variables $\tilde\Psi_m\equiv\sqrt{\tilde\beta(x_m)}\Psi_m$, simplifies to
\begin{equation}\label{flow3}
    \partial_s\tilde\Psi_m=\pm\left(\tilde\beta\partial_x\tilde\Psi_m-\sum_{m,n\in{\mathbb Z}}M_{mn}\tilde\Psi_n \right),
\end{equation}
where $M$ is the infinite-dimensional matrix with $M_{mm}=0$ and
\begin{equation}
    M_{mn}=\frac{\pi}{\beta}\frac{(-1)^{(m-n)\nu_1}\sqrt{\tilde\beta(x_m)\tilde\beta(x_n)}}{\sinh\left[\frac{\pi}{\beta}(x_m-x_{n}-(m-n)L)\right]}
\end{equation}
for $m\ne n$. Note that $M$ is antisymmetric. It is to be regarded as a function of $x$, on which it depends via $\{x_n\}$. Eq.~(\ref{flow3}) is a system of first-order local equations which can be solved by the method of characteristics. The solution is
\begin{equation}
    \tilde\Psi_m(x,s)=\sum_{n\in{\mathbb Z}}O_{mn}^\pm(s)\tilde\Psi_n(x^\pm(s),0),
\end{equation}
where
\begin{equation}\label{curveO}
    x^\pm(s)=z^{-1}(z(x)\pm2\pi s)\qquad O^{\pm}(s)={\cal P}\exp\left[\mp\int_{0}^s ds'M(x^\pm(s'))\right]
\end{equation}
and ${\cal P}$ denotes path ordering. Note that, since $M$ is antisymmetric, $O^\pm$ is an orthogonal matrix. Note also that the curve $x^\pm$ is well-defined because $z$ grows monotonically from $-\infty$ to $\infty$ and hence is invertible. Since the domain of $z$ is the interval $V$, $x^\pm(s)$ lies in $V$ for all values of $s$. Undoing the change of variables and noting that $\Psi_V=\Psi_0$, we obtain
\begin{equation}\label{modev1}
\Psi_V(x,s)=\sum_{n\in{\mathbb Z}}C_{n}^\pm(s)\psi(x_{n}^\pm(s)),
\end{equation}
where
\begin{equation}\label{curvenC}
    x_{n}^\pm(s)=z^{-1}(z(x_n)\pm 2\pi s)\qquad C_{n}^\pm(s)=O_{0n}^\pm(s)\sqrt{\frac{\tilde\beta(x_{n}^\pm(s))}{\tilde\beta(x)}}.
\end{equation}
Thus, the modular evolution of $\psi(x)$ is a linear combination of the values of $\psi$ at infinitely many points of the interval. Both the points and the coefficients of the linear combination depend on the modular time $s$. As is clear from (\ref{curveO}), $O^\pm(0)$ is the identity, so $O_{0n}^\pm(0)=\delta_{0n}$. Moreover, $x_{0}^\pm(0)=x_0=x$, so $\Psi_V(x,0)=\psi(x)$ as it should be. Let us now momentarily attach a superscript $\pm$ to each chirality to distinguish between them, and group them into the Dirac field $\psi_D=(\psi^+,\psi^-)$. According to (\ref{modev1}), the modular evolution $\Psi_{DV}$ of $\psi_D$ is
\begin{equation}\label{modevdirac1}
    \Psi_{DV}(x,s)=\begin{pmatrix}
    \Psi_{V}^+(x,s)\\
    \Psi_{V}^-(x,s)
    \end{pmatrix}=\sum_{n\in{\mathbb Z}}C_{n}(s)\begin{pmatrix}
           \psi^+(x_{n}^+(s)) \\
           \psi^-(x_{n}^-(s))
         \end{pmatrix},
\end{equation}
where
\begin{equation}
    C_{n}(s)=\begin{pmatrix}
         C_{n}^+(s) & 0\\
         0 & C_{n}^-(s)
         \end{pmatrix}.
\end{equation}
Recalling that the Heisenberg evolution $\Psi_D$ of $\psi_D$ is given by $\Psi_D(x,t)=(\psi^+(x+t),\psi^-(x-t))$, we can rewrite (\ref{modevdirac1}) as
\begin{equation}\label{modevdirac2}
    \Psi_{DV}(x,s)=\sum_{n\in{\mathbb Z}}C_{n}(s)\Psi_D(y_n(s),t_n(s)),
\end{equation}
where $y_n(s)\pm t_n(s)=x_{n}^\pm(s)$ and hence
\begin{equation}\label{curves}
    y_n(s)=\frac{x_{n}^+(s)+x_{n}^-(s)}{2}\qquad t_n(s)=\frac{x_{n}^+(s)-x_{n}^-(s)}{2}.
\end{equation}
Note that $(y_n(s),t_n(s))$ is the intersection of two null lines which cross $V$ at $x_{n}^+$ and $x_{n}^-$, and hence it lies in $D(V)$, the causal development of $V$.
Thus, according to (\ref{modevdirac2}), the modular evolution of $\psi_D(x)$ after a modular time $s$ is a linear combination of the values of the Heisenberg Dirac field $\Psi_D$ at a countably infinite set of points in $D(V)$. As $s$ changes, these points move, drawing an infinite set of curves which is represented in Figs.~\ref{fig:diamantes} and 
\ref{fig:diamantes2}.

\section{Entropy}\label{sect:4}

\begin{figure}[t]
     \centering
     \includegraphics[width=\linewidth]{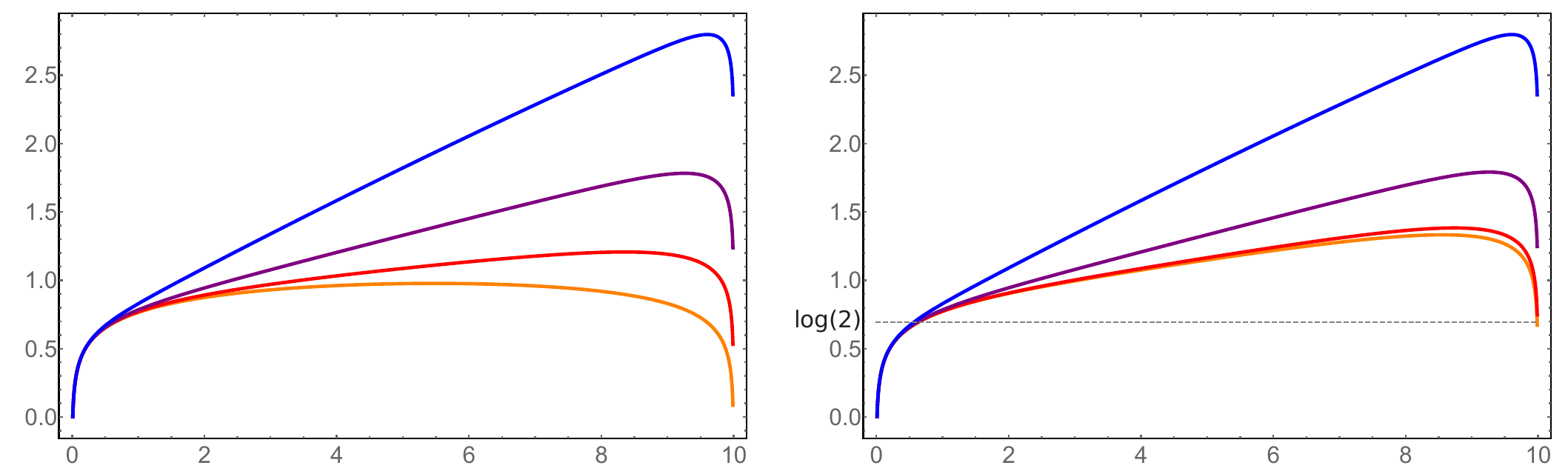} 
   \caption{Entropy for a single interval as a function of its length $\ell$, for different values of $\beta$ ($\beta=2,\,4,\,8,\,16$ from top to bottom) and $L=10$. The left and right panels correspond respectively to Neveu-Schwarz ($\nu_1=1$) and Ramond ($\nu_1=0$) boundary conditions. At high temperatures the entropy is almost linear because thermal effects dominate over entanglement. At zero temperature, the Neveu-Schwarz entropy is symmetric with respect to the point $\ell=L/2$ because the field is in a pure state. Contrarily, the Ramond entropy does not exhibit this behavior because the field is in a mixed state in which the zero mode has $1/2$ probability of being empty or occupied. Correspondingly, the entropy in this case approaches the value $S=\log 2$ as $\ell\to L$.}\label{fig:entropy}
\end{figure}

The resolvent (\ref{resoltorusfinal}) can also be used to compute the entanglement entropy for the region $ V=\bigcup_{i=1}^N(a_i,b_i)$ by substituting it into Eq.~(\ref{EEres}). In doing that, the terms proportional to the identity cancel, and changing the variable of integration from $\xi$ to $q=-k(\xi)$ one finds
\begin{alignat}{2}
    S_V=-i\int_{0}^\infty\frac{dq}{e^{2\pi q}-1}\int_V dx\lim_{y\to x}\frac{1}{\sigma(x-y)}&\left[e^{iq[\omega(x)-\omega(y)]}\frac{\sigma_\nu(x-y-iq\ell)}{\sigma_\nu(-iq\ell)}\right.\nonumber\\
    &\left.-e^{-iq[\omega(x)-\omega(y)]}\frac{\sigma_\nu(x-y+iq\ell)}{\sigma_\nu(iq\ell)}\right],
\end{alignat}
which is independent of the chirality. Since $\sigma(z)\sim z$ near the origin, the limit above is a derivative of the factor between square parentheses,
\begin{equation}\label{entro137}
    S_V=-2i\int_{0}^\infty\frac{dq}{e^{2\pi q}-1}\int_V dx\left\{iq\sum_{i=1}^N\left[\zeta(b_i-x)-\zeta(a_i-x)\right]-\zeta_\nu(iq\ell)\right\},
\end{equation}
where
\begin{equation}
    \zeta_\nu(z)=(\log\sigma_\nu)'(z)=\zeta(z+L/2+i\nu_1\beta/2)-\zeta(L/2)-\nu_1\zeta(i\beta/2).
\end{equation}
Note from the properties of $\zeta$ discussed in appendix \ref{sec:weierstrass} that $\zeta_\nu$ has no poles on the imaginary axis. Note also that it is odd, $\zeta_\nu(-z)=-\zeta_\nu(z)$, and satisfies $\zeta_{\nu}(z^*)=\zeta_{\nu}^*(z)$.
We can see that the $\nu$-dependent term in (\ref{entro137}) is independent of $x$, so the integral over $V$ gives just the total length $\ell$ multiplied by the integrand $\zeta_{\nu}(iq\ell)$. For the other term, the integral in $x$ can be performed introducing a cutoff $\epsilon$ to regulate the divergences coming from the entanglement across the endpoints of the intervals. The result is
\begin{equation}
S_V=\frac{1}{6} \Xi _V + 2i\ell \int_0^{\infty}{ dq \,\frac{\zeta_\nu (iq\ell)}{e^{2\pi q}-1}}\,,\label{finalentropy}
\end{equation}
where we have defined
\begin{equation}\label{letrarara}
    \Xi _V =\sum_{i=1}^N\log\frac{\sigma(b_i-a_i)}{\epsilon}+\sum_{i\ne j}\log\frac{\sigma(|b_i-a_j|)}{\sqrt{\sigma(|a_i-a_j|)\sigma(|b_i-b_j|)}}.
\end{equation}
Note that the dependence on the boundary conditions comes only through the second term in (\ref{finalentropy}), which is real because $\zeta_\nu(iq\ell)$ is imaginary, and well-defined because $\zeta_\nu$ vanishes at the origin and diverges only linearly at infinity. Eqs.~(\ref{finalentropy}) and (\ref{letrarara}) give the entanglement entropy of a chiral fermion on a circle at non-zero temperature, for an arbitrary collection of intervals. This had been computed using the replica trick in \cite{Azeyanagi:2007bj,Herzog:2013py} in the form of a low-temperature and a high-temperature expansion. We believe the above expression is much more practical for applications. 
In Fig.~\ref{fig:entropy} we plot $S_V$ as a function of $\ell$ in the case of a single interval, for different temperatures which increase from bottom to top. 

Let us now explore some limits of the above result. Consider first the limit $L\to\infty$. From (\ref{sigmalim}) and (\ref{zetalim}) we find that, in this limit, the entropy (\ref{finalentropy})-(\ref{letrarara}) takes the form
\begin{alignat}{2}
S_V&=\frac{1}{6}\sum_{i=1}^N\log\left\{\frac{\beta}{\pi\epsilon}\sinh\left[\frac{\pi}{\beta}(b_i-a_i)\right]\right\}\nonumber\\
&+\frac{1}{6}\sum_{i\ne j}\log\frac{\sinh\left(\frac{\pi}{\beta}|b_i-a_j|\right)}{\sqrt{\sinh\left(\frac{\pi}{\beta}|a_i-a_j|\right)\sinh\left(\frac{\pi}{\beta}|b_i-b_j|\right)}},
\end{alignat}
which is independent of the boundary conditions, as it should be.
Further taking the limit $\beta\to\infty$ of the above expression, one recovers the result of \cite{Casini:2009vk} corresponding to the line and the vacuum state. The above expression can also be viewed as the first approximation to the entanglement entropy in the regime $L\gg\beta$. Restricting it to the case of a single interval and letting $\ell\to L$ yields
\begin{equation}
    S=\frac{\pi L}{6\beta},
\end{equation}
as predicted by the Cardy formula. Consider now the limit $\beta\to\infty$, with $L$ finite. From (\ref{sigmalim}) and (\ref{zetalim}) we find that, in this limit, (\ref{finalentropy})-(\ref{letrarara}) reduces to
\begin{alignat}{2}\label{entropyzeroT}
    S_V&=\frac{1}{6}\sum_{i=1}^N\log\left\{\frac{L}{\pi\epsilon}\sin\left[\frac{\pi}{L}(b_i-a_i)\right]\right\}\nonumber\\
&+\frac{1}{6}\sum_{i\ne j}\log\frac{\sin\left(\frac{\pi}{L}|b_i-a_j|\right)}{\sqrt{\sin\left(\frac{\pi}{L}|a_i-a_j|\right)\sin\left(\frac{\pi}{L}|b_i-b_j|\right)}}+g_{\nu_1}(\ell),
\end{alignat}
with
\begin{equation}\label{g}
    g_{0}(\ell)=\frac{2\pi \ell}{L}\int_0^\infty dq\, \frac{\tanh \left( \pi q \ell /L\right)}{e^{2\pi q}-1}\qquad g_1(\ell)=0.
\end{equation}
For Neveu-Schwarz boundary conditions ($\nu_1=1$), the above formula implies $S_V=S_{\bar V}$, where $\bar V$ denotes the complement of $V$,
as corresponds to a pure state which in this case is the vacuum. Contrarily, for Ramond boundary conditions ($\nu_1=0$), the entropy does not have this symmetry property because $g_0(L-\ell)\ne g_0(\ell)$. This is not surprising: the chiral fermion with Ramond boundary conditions has a zero mode which does not contribute to the energy, and therefore its ground state is degenerate; it has degeneracy 2 because the zero mode can be either empty or occupied. Therefore, a Ramond chiral fermion at zero temperature is not in a pure state, but in a mixed state in which the zero mode has $1/2$ probability of being empty or occupied. In fact, in the limit $\ell\to L$ the above formula in the Ramond case gives
\begin{equation}
    S=2\pi\int_{0}^\infty dq\,\frac{\tanh(\pi q)}{e^{2\pi q}-1}=\log 2,
\end{equation}
in agreement with the above discussion. This is also shown in Fig.~\ref{fig:entropy}.

\begin{figure}[t]
     \centering
  \begin{subfigure}[t]{0.5\textwidth}
        \centering
        \includegraphics[width=\linewidth]{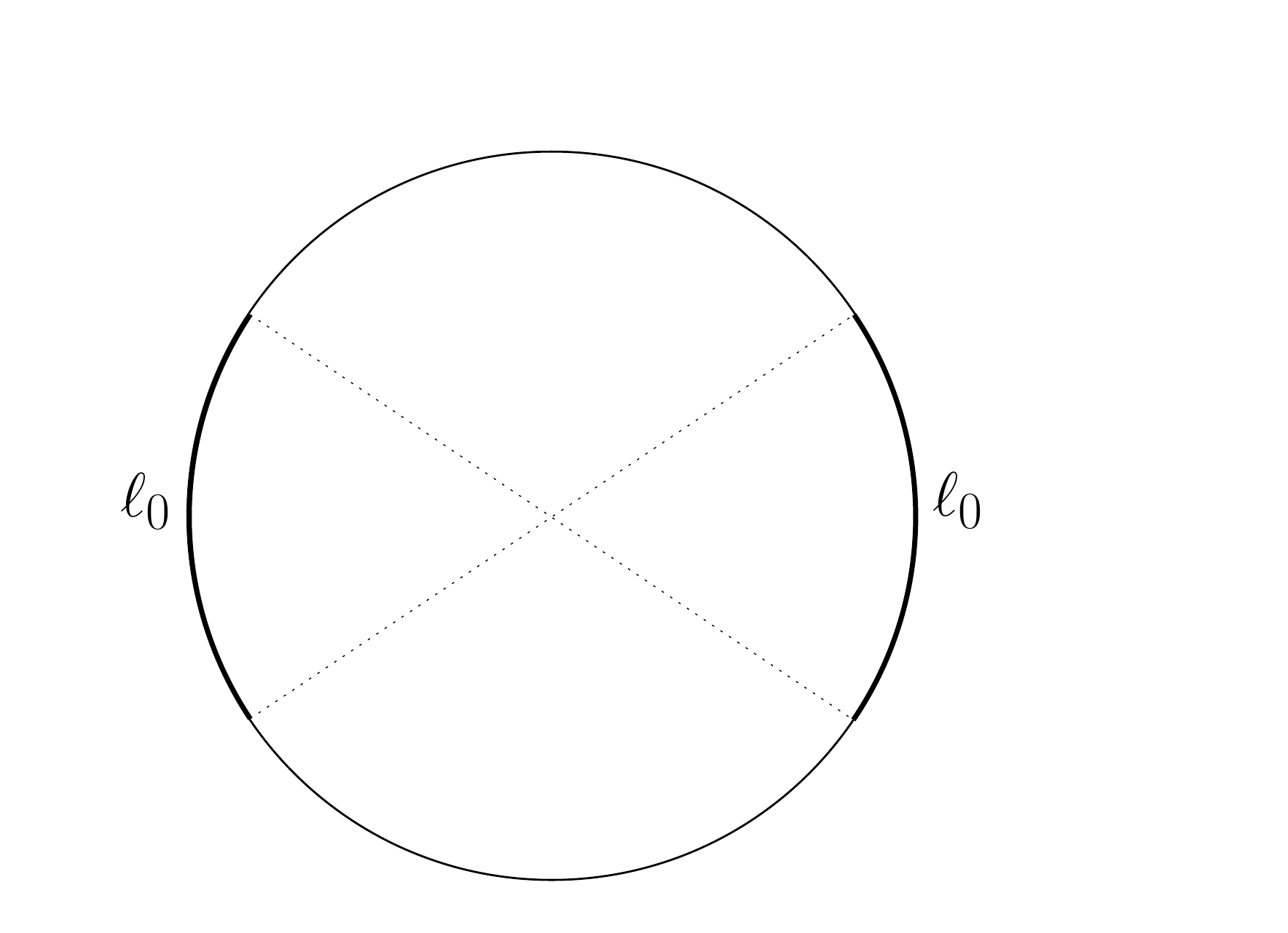} 
      \end{subfigure}
     \begin{subfigure}[t]{0.49\textwidth}
         \centering
        \includegraphics[width=\linewidth]{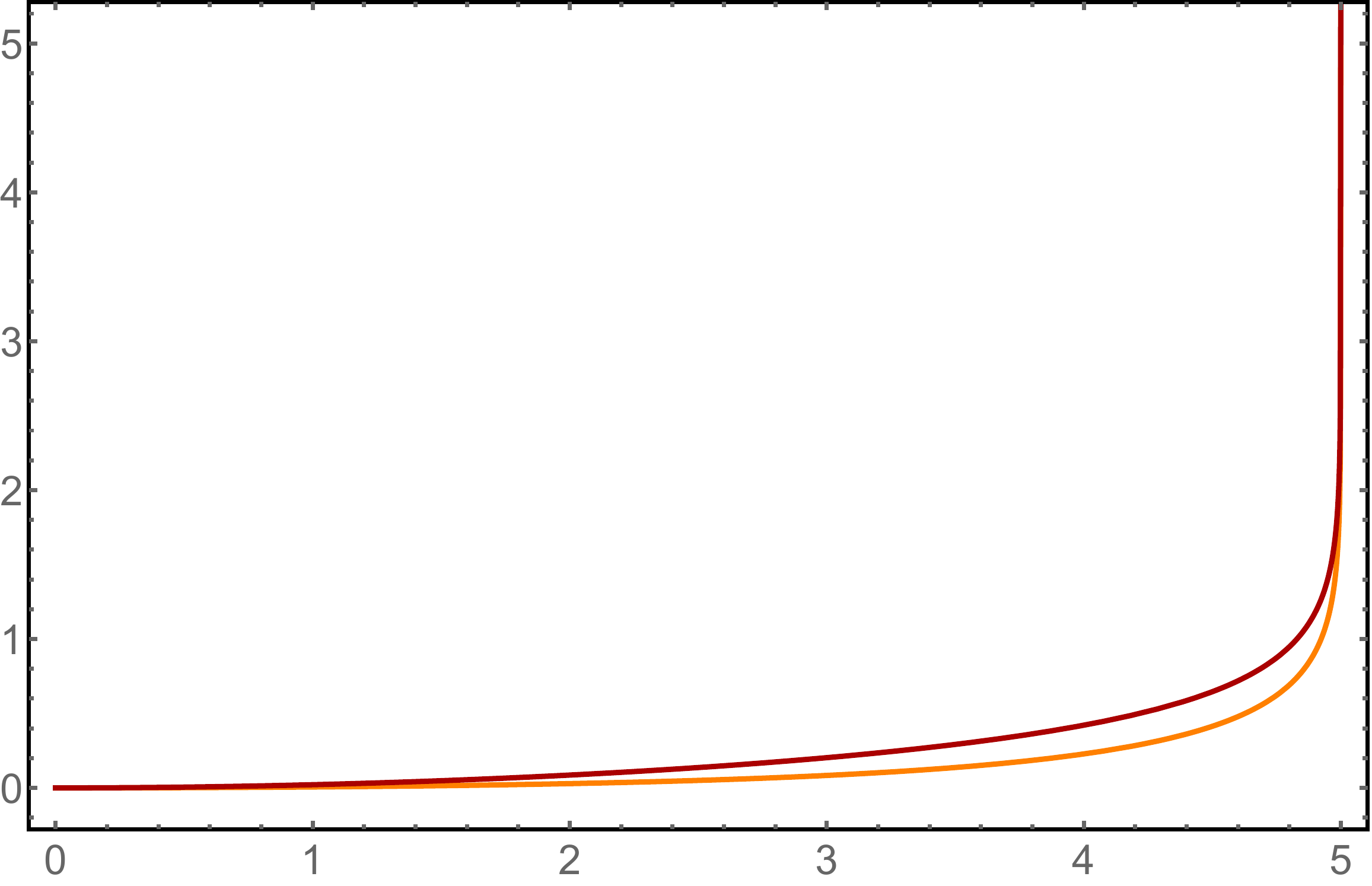} 
             \end{subfigure}
          \caption{Behavior of the mutual information between the two diametrically opposed intervals shown in the figure as a function of their size $\ell_0$, for Neveu-Schwarz (orange) and Ramond (red) boundary conditions. In this plot $L=10$ and $\beta=20$.}\label{fig:mutual}
\end{figure}

We also computed the mutual information between two diametrically opposed intervals of length $\ell_0$. This region and the corresponding mutual information as a function of $\ell_0$ are shown in Fig.~\ref{fig:mutual} (orange curve for Neveu-Schwartz and red curve for Ramond boundary conditions). The mutual information grows as the intervals get larger, which is consistent with the monotonicity property of mutual information. In the limit where the endpoints of the intervals coincide, the mutual information diverges as expected.

\section{Conclusions}\label{sect:5}

In this paper, we have presented the detailed calculation of a new modular Hamiltonian, namely that of a chiral fermion on a circle at non-zero temperature (i.e., on a torus in the Euclidean formalism), for an arbitrary collection of intervals. We have studied the properties of this modular Hamiltonian, checking that it reproduces known results in different limits and analyzing the corresponding modular flow. We have also obtained a simple exact expression for the entanglement entropy, which was previously known only in the form of a low-temperature and a high-temperature expansion \cite{Azeyanagi:2007bj,Herzog:2013py}. 

Our approach is based on a novel and simple application of the method of images, which enables us to map the computation of the resolvent (the key object from which one can derive the modular Hamiltonian and the entanglement entropy) to a similar problem on the plane. As a result, we are able to obtain the resolvent by almost elementary calculations. We believe the method is very powerful and can be applied to other systems, especially free theories such as the scalar field, for which no known results at finite size and finite temperature exist in the literature.

Our result for the modular Hamiltonian is given in Eqs.~(\ref{moddec})-(\ref{nonlocal}). This is a highly non-local object, which couples each point of the region $V$ to an infinite number of points within each interval of $V$, accumulating near the endpoints. The non-locality persists even in the case of a single interval, regardless of the boundary conditions, a behavior that had not been seen in any of the few modular Hamiltonians computed so far. The modular flow inherits this non-locality, and is described by an infinite set of curves which, as guaranteed by the Tomita-Takesaki theory, are contained in the causal development of $V$.

The entanglement entropy is given in Eqs.~(\ref{finalentropy})-(\ref{letrarara}) for an arbitrary collection of intervals, and its behavior for a single interval is plotted in Fig.~\ref{fig:entropy}. In the Neveu-Schwarz case, the entropy of $V$ becomes equal to that of its complement when the temperature tends to zero, consistently with the fact that, in that limit, the global state tends to a pure state (the vacuum). Interestingly, in the Ramond case the entropy does not exhibit this behavior, because the zero-temperature state is a mixed state in which the zero mode has $1/2$ probability of being empty or occupied.

It is interesting to note that, for chiral theories, which suffer from a gravitational anomaly, the entanglement entropy is anomalous: two regions with the same domain of dependence have different entanglement entropies \cite{Nishioka:2015uka,Iqbal:2015vka,Hughes:2015ora}. In this paper we have not seen this explicitly because we have considered only subsets of the $t=0$ circle (so we cannot compare between two regions with the same domain of dependence), but the generalization of our results to subsets of an arbitrary Cauchy surface is straightforward, and one can see that the anomaly is indeed present (even in the simplest case of a line at zero temperature \cite{Casini:2009vk}).
The anomaly, however, disappears when the entropies of both chiralities are added to obtain the entropy of the Dirac fermion.

As mentioned above, we expect the applications of the method of images to go beyond the fermion case; we are currently following this strategy to study the entanglement properties of a scalar field on the torus. On the other hand, the resolvent we computed can find other applications beyond those presented here.
For example, one can use it to study the effect of adding a small mass along the lines of \cite{Casini:2009vk}. Some features of the entanglement and Renyi entropies for a massive fermion on the torus have been studied before in \cite{Herzog:2013py}, but we believe that more analytical progress can be achieved using the results of this paper.

\acknowledgments

The authors thank Horacio Casini, Gaston Giribet, Andr\'es Goya and Mart\'in Mereb for useful discussions. This work has been supported by UBA and CONICET. A.~G.~ would like to acknowledge the kind hospitality of the ICTP during the last stages of completion of the present work. 

\appendix

\section{The Weierstrass functions}
\label{sec:weierstrass}

In this appendix we briefly review the Weierstrass functions, see e.g.~\cite{chandrasekharan2012elliptic,Pastras:2017wot} for more details. We start with some generalities about elliptic functions. A function $f:{\mathbb C}\to{\mathbb C}$ is said to be elliptic if it is meromorphic (i.e., analytic except for a discrete set of poles) and doubly periodic, which means that 
there exist non-collinear complex numbers $P_1$ and $P_2$ such that
\begin{equation}
    f(z+P_i)=f(z).
\end{equation}
The periods $P_1$ and $P_2$ are said to be fundamental if the parallelogram they define (which by convention includes the vertex at the origin and the two edges emanating from it, but not the other edges and vertices) contains no point $P\ne 0$ such that $f(z+P)=f(z)$. In that case, the parallelogram is called a fundamental period parallelogram, and any translation of it is called a cell.
Two important properties of elliptic functions are: within a cell, (i) the sum of all residues vanishes, and (ii) the total number of poles is equal to the total number of zeros (each pole and zero weighted by its multiplicity). This number is independent of the cell, and it is called the order of the function. Note that, by property (i), there is no elliptic function of order 1. On the other hand, Liouville's theorem implies that an elliptic function of order 0 is necessarily constant. Therefore, the lowest possible order of a non-constant elliptic function is 2.

\begin{figure}[t]
    \centering
    \includegraphics[width=0.7\linewidth]{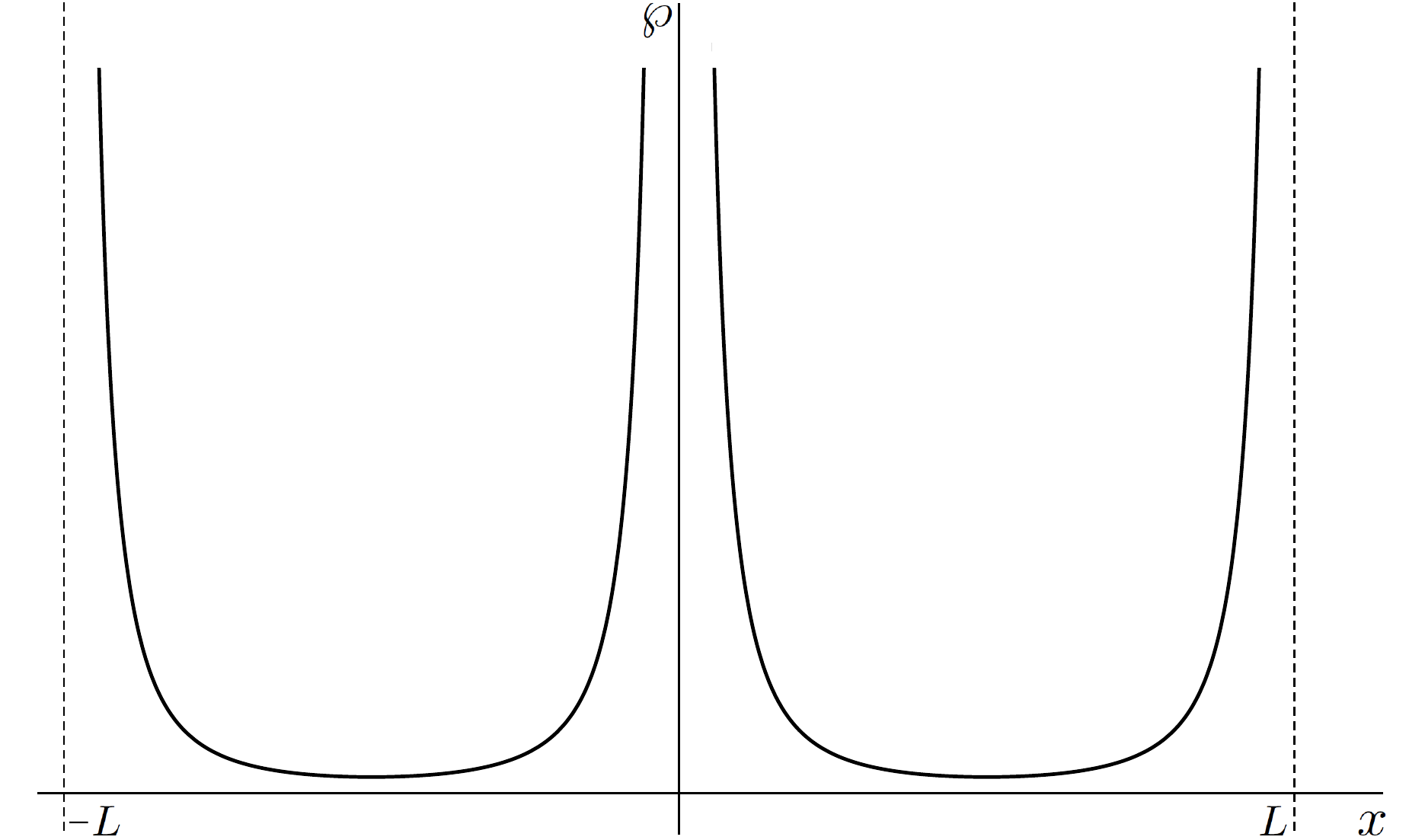} 
    \caption{The Weierstrass $\wp$ function with $P_1=L, P_2=i\beta$, evaluated on the real line.} \label{fig:p}
\end{figure}

An example of an elliptic function of order 2 is the Weierstrass elliptic function. Let $P_1$ and $P_2$ be non-collinear complex numbers. The Weierstrass elliptic function $\wp$ associated with the lattice $\Lambda=\{\lambda_1 P_1+\lambda_2 P_2,\lambda_i\in{\mathbb Z}\}$ is defined by
\begin{equation}\label{wp}
     \wp(z)=\frac{1}{z^2}+\sum_{\lambda\ne 0}\left[\frac{1}{(z+\lambda)^2}-\frac{1}{\lambda^2}\right],
\end{equation}
where $\lambda$ runs over $\Lambda$. One can easily check that $\wp$ is indeed elliptic with fundamental periods $P_1$ and $P_2$. Within the fundamental period parallelogram, it has a double pole at the origin and no other poles, so it is of order 2.
Note that it is even, $\wp(-z)=\wp(z)$. Together with periodicity, this implies that it is also symmetric, $\wp(\omega_i-z)=\wp(\omega_i+z)$,  with respect to the points $\omega_1=P_1/2$, $\omega_2=P_2/2$ and $\omega_3=(P_1+P_2)/2$, all lying in the fundamental period parallelogram. Since these three points are not poles, it follows that they are extrema of $\wp$. In fact, they are the only extrema of $\wp$ within the fundamental period parallelogram, because $\wp'$ is an elliptic function of order 3 and hence has three zeros in each cell.
Computing the Laurent expansion of $\wp$ around the origin and defining
\begin{equation}
    g_2=60\sum_{\lambda\ne 0}\frac{1}{\lambda^4}\qquad g_3=140\sum_{\lambda\ne 0}\frac{1}{\lambda^6},
\end{equation}
one finds that the function $(\wp')^2-4\wp^3+g_2\wp+g_3$ is analytic and vanishes at the origin. Since this function is elliptic, it must vanish everywhere, so that
\begin{equation}\label{weierstrassdiff}
    [\wp'(z)]^2=Q(\wp(z))\qquad Q(t)=4t^3-g_2 t-g_3.
\end{equation}
This is the Weierstrass differential equation. Note that it implies that $\wp(\omega_1)$, $\wp(\omega_2)$ and $\wp(\omega_3)$ are the three solutions of the cubic equation $Q(t)=0$. In other words,
\begin{equation}\label{polynomial}
    Q(t)=4[t-\wp(\omega_1)]]t-\wp(\omega_2)][t-\wp(\omega_3)]
\end{equation}
for all $t\in{\mathbb C}$. Equating the coefficients of each power of $t$, one obtains
\begin{alignat}{2}\label{rootrelations}
    &\wp(\omega_1)+\wp(\omega_2)+\wp(\omega_3)=0\nonumber\\
    &\wp(\omega_1)\wp(\omega_2)+\wp(\omega_2)\wp(\omega_3)+\wp(\omega_1)\wp(\omega_3)=-\frac{g_2}{4}\nonumber\\
    &\wp(\omega_1)\wp(\omega_2)\wp(\omega_3)=\frac{g_3}{4}.
\end{alignat}
On the other hand, differentiating (\ref{weierstrassdiff}) with respect to $z$ and using that $\omega_1$, $\omega_2$ and $\omega_3$ are necessarily first-order roots of $\wp'$, one finds that $\wp(\omega_1)$
$\wp(\omega_2)$ and $\wp(\omega_3)$ are first-order roots of $Q$, and hence they are all different from each other. Now, the case of interest for us is $P_1=L$, $P_2=i\beta$.
In this case the lattice $\Lambda$ is invariant under complex conjugation, from which it follows that
\begin{equation}\label{real}
    \wp(z^*)=\wp^*(z),
\end{equation}
so $\wp$ is real when evaluated on the real line. Let us study the behavior of $\wp$ on the interval $(0,L)$, which determines its behavior throughout the real line by periodicity. As we have seen, $\wp$ is symmetric with respect to the point $\omega_1=L/2$, where it has an extremum. This is the only extremum within the interval, because the other two extrema in the fundamental period parallelogram, $\omega_2$ and $\omega_3$, are not real. On the other hand, it is clear that $\wp(x)\to\infty$ as $x\to 0,L$, so $x=L/2$ is the absolute minimum of $\wp$ within the interval. Furthermore, $\wp(L/2)>0$, so $\wp$ is positive. To see this,
note first that, as a consequence of (\ref{real}) and the periodicity of $\wp$, the three roots $\wp(\omega_1)$, $\wp(\omega_2)$ and $\wp(\omega_3)$ of the cubic polynomial $Q$ are all real. 
Moreover, for $x\in(0,L/2)$ we have $\wp(x)\in(\wp(\omega_1),\infty)$ and $\wp'(x)\ne 0$, so, by (\ref{weierstrassdiff}), $Q$ has no roots in the interval $(\wp(\omega_1),\infty)$. Therefore, $\wp(\omega_1)$ is the largest root. Since the three roots are all different from each other, the first equation in (\ref{rootrelations}) implies $\wp(\omega_1)>0$, as we wanted to show. The behavior of $\wp$ on the real line is shown in Fig.~\ref{fig:p}.

\begin{figure}[t]
    \centering
    \includegraphics[width=\linewidth]{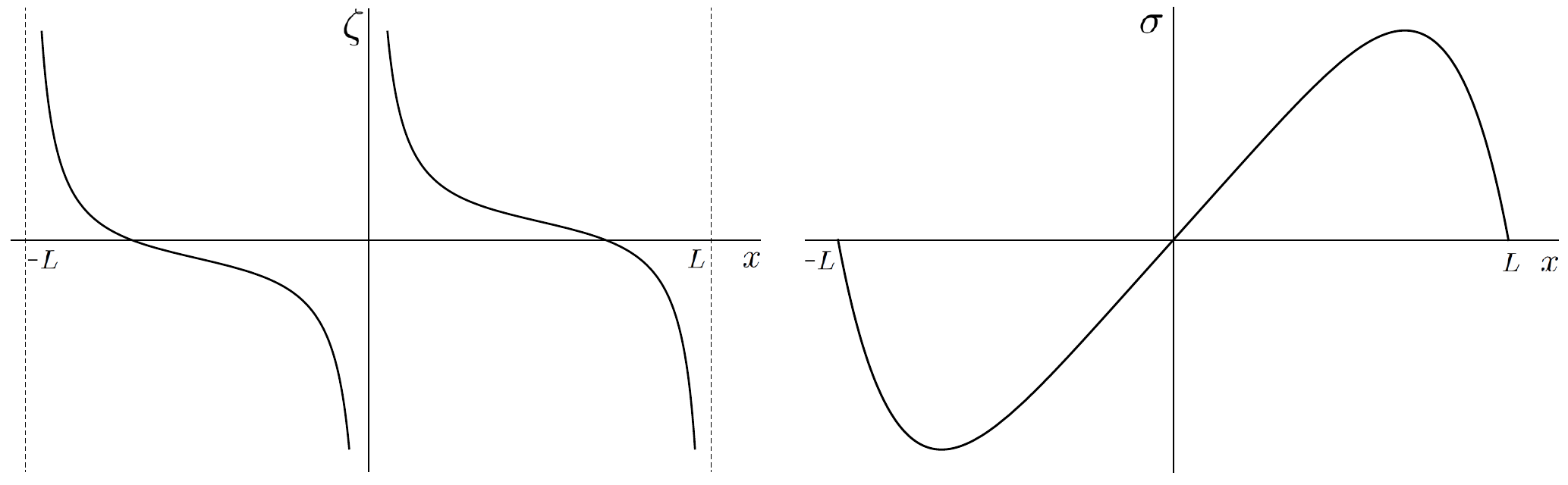} 
    \caption{The Weierstrass $\zeta$ and $\sigma$ functions with $P_1=L, P_2=i\beta$, evaluated on the real line.} \label{fig:zetasigma}
\end{figure}

The Weierstrass elliptic function has associated two quasiperiodic functions called the Weierstrass zeta and sigma functions.
The Weierstrass zeta function $\zeta$ is defined by
\begin{equation}
    \zeta(z)=\frac{1}{z}+\sum_{\lambda\ne 0}\left(\frac{1}{z+\lambda}-\frac{1}{\lambda}+\frac{z}{\lambda^2}\right).
\end{equation}
Clearly, this is an odd function satisfying $\zeta'=-\wp$. The latter property, together with the periodicity of $\wp$, implies $\zeta(z+P_i)=\zeta(z)+c$ for some constant $c$. Evaluating this equation at $z=-P_i/2$ and using that $\zeta$ is odd yields $c=2\zeta(P_i/2)$, so that
\begin{equation}\label{zetap}
    \zeta(z+P_i)=\zeta(z)+2\zeta(P_i/2).
\end{equation}
Note also that the poles of $\zeta$ are simple with unit residue, and they lie at the points congruent to the origin (i.e., which differ from the origin by an element of the lattice $\Lambda$).
Integrating $\zeta$ along the boundary of a cell and using the above quasiperiodicity property, one obtains
\begin{equation}\label{halfperiods}
    P_2\zeta(P_1/2)-P_1\zeta(P_2/2)=i\pi
\end{equation}
provided that the ratio $P_2/P_1$ has a pòsitive imaginary part (in the opposite case there is a minus sign on the right-hand side above). The Weierstrass sigma function $\sigma$ is defined by
\begin{equation}\label{sigma}
    \sigma(z)=z\prod_{\lambda\ne 0}\left[\left(1+\frac{z}{\lambda}\right)e^{-\frac{z}{\lambda}+\frac{1}{2}\left(\frac{z}{\lambda}\right)^2}\right].
\end{equation}
This is an odd function satisfying $\sigma'/\sigma=\zeta$. The latter property, together with (\ref{zetap}), implies $\sigma(z+P_i)=Ce^{2\zeta(P_i/2)z}\sigma(z)$ for some constant $C$. Evaluating this equation at $z=-P_i/2$ and using that $\sigma$ is odd yields $C=-e^{\zeta(P_i/2)P_i}$, so that
\begin{equation}\label{sigmap}
    \sigma(z+P_i)=-e^{\zeta(P_i/2)(2z+P_i)}\sigma(z).
\end{equation}
Note also that $\sigma$ is analytic, has a simple zero at each point congruent to the origin and is non-vanishing everywhere else and satisfies $\sigma'(0)=1$. In the case of interest for us, $P_1=L$, $P_2=i\beta$, the Weierstrass zeta and sigma functions satisfy $f(z^*)=f^*(z)$, like the Weierstrass elliptic function, so they are real when evaluated on the real line. We have $\zeta(x)\to\infty$ as $x\to 0^+$ and $\zeta(x)\to-\infty$ as $x\to L^-$. Moreover, $\zeta$ is monotonically decreasing in the interval $(0,L)$ due to the positivity of $\wp$. On the other hand, $\sigma(x)=0$ at $x=0,L$ and $\sigma(x)\ne 0$ for $x\in(0,L)$. Since $\sigma'(0)=1$, it follows that $\sigma$ is positive within the interval $(0,L)$. The behavior of $\zeta$ and $\sigma$ on the real line is shown in Fig.~\ref{fig:zetasigma}.

Let us finally study the behavior of the Weierstrass functions as one of the periods, say $P_j$, goes to infinity while the other period, say $P_i$, remains finite. In that case, all factors with $\lambda_j\ne 0$ in (\ref{sigma}) are equal to 1, so we can write
\begin{alignat}{2}\label{sigmalim}
\sigma(z)&=z\prod_{n\ne 0}\left[\left(1+\frac{z}{nP_i}\right)e^{-\frac{z}{nP_i}+\frac{1}{2}\left(\frac{z}{nP_i}\right)^2}\right]\nonumber\\
&=z\prod_{n=1}^\infty\left[1-\left(\frac{z}{nP_i}\right)^2\right]e^{\left(\frac{z}{P_i}\right)^2\sum_{n=1}^\infty n^{-2}}\nonumber\\
&=\frac{P_i}{\pi}\sin\left(\frac{\pi z}{P_i}\right)e^{\frac{1}{6}\left(\frac{\pi z}{P_i}\right)^2}.
\end{alignat}
In the last step we have used the Euler product formula for the sine and recognized the Riemann zeta function $\zeta(2)=\pi^2/6$ in the exponent. From this expression we obtain
\begin{alignat}{2}\label{zetalim}
    &\zeta(z)=\frac{\pi}{P_i}\left[\cot\left(\frac{\pi z}{P_i}\right)+\frac{1}{3}\frac{\pi z}{P_i}\right]\nonumber\\
    &\wp(z)=\left(\frac{\pi}{P_i}\right)^2\left[\frac{1}{\sin^2\left(\frac{\pi z}{P_i}\right)}-\frac{1}{3}\right].
\end{alignat}
One can check that, in the case $P_1=L$, $P_2=i\beta$, these limiting expressions have the qualitative features discussed above for generic values of $L$ and $\beta$.

\section{Monotonicity of $z$}
\label{sec:monotonicity}

In the previous appendix we have shown that the Weierstrass zeta function is monotonically decreasing between any two consecutive poles in the real line. In fact, it also satisfies
\begin{equation}\label{property}
    \zeta(x)-\zeta(y)>\frac{2\zeta(L/2)}{L}(x-y)
\end{equation}
provided that $x<y$ and there is no pole of $\zeta$ between $x$ and $y$.
This can be seen by using the relation between the Weierstrass elliptic function and the Jacobi theta function $\theta_1$,
\begin{equation}\label{ptheta}
    \wp(x)=-\frac{2\zeta(L/2)}{L}-\left(\frac{\pi}{L}\right)^2(\log\theta_1)''(\pi x/L),
\end{equation}
and also the expansion
\begin{equation}
    (\log\theta_1)'(x)=\cot x+4\sin(2x)\sum_{n=1}^{\infty}\frac{q^{2n}}{1-2q^{2n}\cos(2x)+q^{4n}},
\end{equation}
where $q=e^{-\pi\beta/L}$.
Indeed, from the above expansion one sees that $(\log\theta_1)''(\pi/2)<0$, which, together with (\ref{ptheta}), implies $\wp(L/2)>-2\zeta(L/2)/L$. Since $x=L/2$ is the absolute minimum of $\wp$, we have $\wp(x)>-2\zeta(L/2)/L$ for all $x\in{\mathbb R}$, and integrating this relation yields (\ref{property}). Let us now use this property to show that the function $z$ defined in Eq.~(\ref{z}) is monotonically increasing on each interval of $V$, i.e., that $z'>0$. We have
\begin{alignat}{2}
    z'(x)&=\sum_{i=1}^N\left[\zeta(b_i-x)-\zeta(a_i-x)\right]-\frac{2\ell}{i\beta}\zeta(i\beta/2)\nonumber\\
    &>\sum_{i=1}^N\left[\zeta(b_i-x)-\zeta(a_i-x)\right]-\frac{2\ell}{L}\zeta(L/2)\equiv f(x),
\end{alignat}
where we have used the relation (\ref{halfperiods}) between the values of $\zeta$ at the half-periods. Rearranging the sum and using the quasiperiodicity of $\zeta$, Eq.~(\ref{zetap}), we can rewrite $f$ as
\begin{alignat}{2}
    f(x)&=\sum_{i=1}^{N-1}\left[\zeta(b_i-x)-\zeta(a_{i+1}-x)\right]+\zeta(b_N-x)-\zeta(a_1-x)-\frac{2\ell}{L}\zeta(L/2)\nonumber\\
    &=\sum_{i=1}^{N-1}\left[\zeta(b_i-x)-\zeta(a_{i+1}-x)\right]+\zeta(b_N-x)-\zeta(a_1-x+L)-\frac{2\zeta(L/2)}{L}(\ell-L).
\end{alignat}
For each difference of zeta functions in the second line above, the arguments satisfy the conditions of (\ref{property}). Using that property one immediately sees that $f>0$, and hence $z'>0$, as we wanted to show.

\bibliography{references}
\bibliographystyle{JHEP}

\end{document}